\documentclass[usenatbib]{mn2e}
\usepackage{graphicx,natbib}
\usepackage{bm,url}
\usepackage{amsmath}
\usepackage{amssymb}
\usepackage{graphicx}
\usepackage{euscript}
\graphicspath{{./figures/}{./png/}}
\newcommand{\EQ}{\begin{equation}}
\newcommand{\EN}{\end{equation}}
\newcommand{\EQA}{\begin{eqnarray}}
\newcommand{\ENA}{\end{eqnarray}}

\newcommand{\Eq}[1]{Eq.~(\ref{#1})}
\newcommand{\Eqs}[2]{Eqs.~(\ref{#1}) and~(\ref{#2})}

\newcommand{\Sec}[1]{Sect.~\ref{#1}}
\newcommand{\Secs}[2]{Sects.~\ref{#1} and \ref{#2}}
\newcommand{\Fig}[1]{Fig.~\ref{#1}}

\newcommand{\Figs}[2]{Figs.~\ref{#1} and \ref{#2}}
\newcommand{\Figss}[2]{Figs.~\ref{#1}--\ref{#2}}
\newcommand{\Tab}[1]{Table~\ref{#1}}
\newcommand{\Tabs}[2]{Tables~\ref{#1} and \ref{#2}}

{}
{}
{}

{}
{}
{}
{}
{}
{}
{}
{}
{}
{}
{}
{}
{}
{}
{}
{}
{}
{}
{}
{}
{}
{}

{}
{}
{}

{}

{}
{}

\newcommand{\BB}{\bm{B}}

\newcommand{\JJ}{\mbox{\boldmath $J$} {}}

\newcommand{\ff}{\mbox{\boldmath $f$} {}}

\newcommand{\const}{{\rm const}  {}}

\def\ga{\mathrel{\mathchoice {\vcenter{\offinterlineskip\halign{\hfil
$\displaystyle##$\hfil\cr>\cr\sim\cr}}}
{\vcenter{\offinterlineskip\halign{\hfil$\textstyle##$\hfil\cr>\cr\sim\cr}}}
{\vcenter{\offinterlineskip\halign{\hfil$\scriptstyle##$\hfil\cr>\cr\sim\cr}}}
{\vcenter{\offinterlineskip\halign{\hfil$\scriptscriptstyle##$\hfil\cr>\cr\sim\cr}}}}}
\def\Ma{\mbox{\rm Ma}}

\def\cs{c_{\rm s}}

\def\Hp{H_{p}}

\def\vA{v_{\rm A}}

\def\lf{l_{\rm f}}

\def\kf{k_{\rm f}}
\def\urms{u_{\rm rms}}

\def\nut{\nu_{\rm t}}

\def\half{{\textstyle{1\over2}}}

\def\onethird{{\textstyle{1\over3}}}

\newcommand{\G}{\,{\rm G}}

\newcommand{\uHz}{\,\mu{\rm Hz}}

\newcommand{\kmeter}{\,{\rm km}}

\newcommand{\Mm}{\,{\rm Mm}}

\newcommand{\yapj}[3]{ #1, {ApJ,} {#2}, #3}

\newcommand{\yapjl}[3]{ #1, {ApJ,} {#2}, #3}

\newcommand{\yana}[3]{ #1, {A\&A,} {#2}, #3}

\newcommand{\ypf}[3]{ #1, {Phys.\ Fluids,} {#2}, #3}

\newcommand{\yaraa}[3]{ #1, {ARA\&A,} {#2}, #3}

\newcommand{\ymn}[3]{ #1, {MNRAS,} {#2}, #3}
\newcommand{\ynat}[3]{ #1, {Nature,} {#2}, #3}
\newcommand{\ysci}[3]{ #1, {Science,} {#2}, #3}
\newcommand{\ysph}[3]{ #1, {Solar Phys.,} {#2}, #3}

\newcommand{\ypnas}[3]{ #1, {Proc.\ Nat.\ Acad.\ Sci.,} {#2}, #3}

\newcommand{\yjour}[4]{ #1, {#2}, {#3}, #4}

\newcommand{\ybook}[3]{ #1, {#2} (#3)}
\newcommand{\yproc}[5]{ #1, in {#3}, ed.\ #4 (#5), #2}

\newcommand{\beq}{\begin{equation}}
\newcommand{\eeq}{\end{equation}}

\newcommand{\ex}{\mbox{{\boldmath $e$}}_{x}}

\newcommand{\ez}{\mbox{{\boldmath $e$}}_{z}}

\newcommand{\Ratio}{q}

\newcommand{\re}{{\rm Re}}
\newcommand{\rc}{{\rm Rc}}

\newcommand{\md}{{\rm Ma}}
\newcommand{\ul}{u_{\rm rms,d}}

\newcommand{\Lu}{L_{z\rm u}}
\newcommand{\Ld}{L_{z\rm d}}
\newcommand{\ru}{{\rho_{\rm u}}}
\newcommand{\rd}{{\rho_{\rm d}}}
\newcommand{\vaxu}{v_{{\rm A}x\rm u}}
\newcommand{\vaxd}{v_{{\rm A}x\rm d}}

\newcommand{\vazd}{v_{{\rm A}z\rm d}}
\newcommand{\csu}{c_{\rm su}}
\newcommand{\csd}{c_{\rm sd}}
\newcommand{\vau}{v_{\rm Au}}
\newcommand{\vad}{v_{\rm Ad}}
\newcommand{\ofn}{{\omega_{{\rm f}\#}}}
\newcommand{\oofn}{{\omega_{{\rm f}\#}^2}}
\newcommand{\ofmn}{{\omega_{{\rm fm}\#}}}
\newcommand{\oofmn}{{\omega_{{\rm fm}\#}^2}}
\newcommand{\kfm}{{k_x^{f\rm max}}}
\newcommand{\Rsun}{R_\odot}

\title{Properties of $p\,$- and $f$-modes in hydromagnetic turbulence}
\author[N. K. Singh et al.]{Nishant K. Singh$^1$\thanks{E-mail:nishant@nordita.org},
Axel Brandenburg$^{1,2}$, S. M. Chitre$^3$, and Matthias Rheinhardt$^4$
\\
$^1$Nordita, KTH Royal Institute of Technology and Stockholm University,
Roslagstullsbacken 23, SE-10691 Stockholm, Sweden\\
$^2$Department of Astronomy, Stockholm University, SE-10691 Stockholm, Sweden\\
$^3$Centre for Basic Sciences, University of Mumbai, Mumbai 400098, India\\
$^4$Physics Department, Gustaf H\"allstr\"omin katu 2a, PO Box 64,
FI-00014 University of Helsinki, Finland
}
\date{\today,~ $ $Revision: 1.307 $ $}

\begin{document}

\maketitle

\begin{abstract}
With the ultimate aim of using the fundamental or $f$-mode to study
helioseismic aspects of turbulence-generated magnetic flux concentrations,
we use randomly forced hydromagnetic simulations of a piecewise isothermal
layer in two dimensions with reflecting boundaries at top and bottom.
We compute numerically diagnostic wavenumber--frequency diagrams of the
vertical velocity at the interface between the denser gas below and the
less dense gas above.
For an Alfv{\'e}n-to-sound speed ratio of about 0.1, a 5\% frequency increase
of the $f$-mode can be measured when $k_x\Hp=3$--$4$, where $k_x$ is the
horizontal wavenumber and $\Hp$ is the pressure scale height at the surface.
Since the solar radius is about 2000 times larger than $\Hp$, the corresponding
spherical harmonic degree would be 6000--8000.
For weaker fields, a $k_x$-dependent frequency decrease by the turbulent
motions becomes dominant.
For vertical magnetic fields, the frequency is enhanced for $k_x\Hp\approx4$,
but decreased relative to its nonmagnetic value for $k_x\Hp\approx9$.
\end{abstract}


\section{Introduction}

Much of our knowledge of the physics beneath the solar photosphere
is obtained from theoretical calculations and simulations.
Helioseismology provides a window to measure certain properties
inside the Sun; see the review by \cite{GBS10}.
This  technique uses sound waves ($p\,$-modes) and to some extent
surface gravity waves ($f$-modes), but the presence of
magnetic fields gives rise to
magneto-acoustic and magneto-gravity waves, whose restoring forces are
caused by magnetic fields modified by pressure and buoyancy forces
\citep[see, e.g.,][]{T83,C11}.
This complicates their use in helioseismology, where magnetic fields are
often not fully self-consistently included.
This can lead to major uncertainties.

Recent detections of changes in the sound travel time at a depth
of some 60\,Mm beneath the surface about 12 hours prior to the emergence
of a sunspot \citep{Ilo11} have not been verified by other groups
\citep{Braun12,Birch13}.
Also the recent proposal of extremely low flow speeds of the
supergranulation \citep{Hanasoge} is in stark contrast to
our theoretical understanding and poses serious challenges.
It is therefore of interest to use simulations to explore theoretically how
such controversial results can be understood; see, e.g., \cite{Geo07}
and \cite{KKMW11} for earlier attempts trying to construct synthetic
helioseismic data from simulations.

The ultimate goal of our present study is to explore the possibility
of using numerical simulations of forced turbulence to assess the effects
of subsurface magnetic fields on the $p\,$- and $f$-modes.
Subsurface magnetic fields can have a broad range of origins.    
The most popular one is the buoyant rise and emergence of flux tubes
deeply rooted at or even below the base of the convection zone \citep{Cal95}.
Another proposal is that global magnetic fields are generated in the
bulk of the convection zone with equatorward migration being
promoted by the near-surface shear layer \citep{B05}.
In this case, subsurface magnetic fields are expected to be concentrated into
sunspots through local effects such as supergranulation \citep{SN12}
or through downflows caused by negative effective magnetic pressure
instability \citep{BKR13,BGJKR14}.
The latter mechanism requires only stratified turbulence and its
operation can be demonstrated and studied in isolation from other
effects using just an isothermal layer.
To examine seismic effects of magnetic fields on the $f$-mode, one must
however introduce a sharp density drop, which implies a corresponding
temperature increase.
This leads us to studying a {\em piecewise} isothermal layer.

In the Sun, waves are excited by convective motions \citep{Ste67,GK88,GK90}.
However, in an isothermal layer there is no convection and turbulence
must be driven by external forcing, as it has been done extensively
in the study of negative effective magnetic pressure effects.
We adopt this method also in the present work, but
use a rather low forcing amplitude to minimize the 
nonlinear effects of large Reynolds numbers and
Mach numbers close to unity.

In the absence of a magnetic field, linear perturbation theory gives
simple expressions for the dispersion relations of $p\,$- and $f$-modes,
which are also the modes which we focus on in this work. 
For large horizontal wavenumbers $k_{\rm h}$, the frequencies
of the solar $f$-mode have been observed to be
significantly smaller than the theoretical estimates, and
both line shift and line width grow with $k_{\rm h}$ \citep{FSTT92,DKM98}.
Both effects are expected to arise due to turbulent background
motions \citep{MR93a,MR93b,MMR99,M00a,M00b,MKE08}.
There have also been alternative proposals to explain 
the frequency shifts as being due to what 
\cite{RG94} call an interfacial wave that depends crucially
on the density stratification of the transition region
between chromosphere and corona; see also \cite{RC95}.

In the presence of magnetic fields, both $p\,$- and $f$-mode frequencies
are affected.
\cite{NT76} derive the dispersion relation for sound waves in the
uniformly horizontally magnetized isothermally stratified half-space
with a rigid lower boundary.
In the presence of structured magnetic fields, e.g., near sunspots,
a process called mode conversion can occur,
i.e., an exchange of energy between fast and slow magnetosonic modes,
which leads to $p\,$-mode absorption in sunspots \citep{Cal06,SC06}.
The properties of surface waves in magnetized
atmospheres have been studied in detail \citep{R81,MR89,MR92,MAR92}.
From these results, one should 
expect changes in the $f$-mode during the course of the solar cycle.
Such variations 
have indeed been observed \citep{ABPP00,DG05} and may be caused by subsurface
magnetic field variations.
It was argued by \cite{DGS01} that the time-variation of $f$-mode
frequencies could be attributed to the presence of a perturbing 
magnetic field of order $20\G$ localized in the outer 1\% of the
solar radius.
%\cite{SKGD97} and \cite{Ant98} deduced from accurately measured
%$f$-mode frequencies the seismic radius of the Sun.
%They found that the customarily accepted value of the solar radius,
%$695.99\Mm$ needs to be reduced by about $200$--$300\kmeter$ to have an
%agreement with the observed $f$-mode frequencies.

The temporal variation of $f$-mode frequencies may be resolved into two
components: an oscillatory component with a one-year period which is
probably an artifact of data analysis resulting from the orbital period
of the Earth, and another slowly varying secular component which appears to
be correlated with the solar activity cycle.
Subsequent work by \cite{Ant03} showed that
variations in the thermal structure of the Sun tend to cause much smaller
shifts in $f$-mode frequencies as compared to those in $p\,$-mode frequencies and as
such are not effective in accounting for the observed $f$-mode variations.
%AB: new paragraph

%AB: new paragraph, and move the 5 last lines from 2 paras above to here
\cite{SKGD97} and \cite{Ant98} deduced from accurately measured
$f$-mode frequencies the seismic radius of the Sun.
They found that the customarily accepted value of the solar radius,
$695.99\Mm$ needs to be reduced by about $200$--$300\kmeter$ to have an
agreement with the observed $f$-mode frequencies.
%AB: continue with previous stuff
From the study of temporal variations of $f$-mode frequencies, \cite{LK05}
%found evidence for time variations of the solar radius
%with an oscillation in antiphase with the solar cycle above $0.99\Rsun$,
%AB: modified, as also now done in proofs
found evidence for time variations of the seismic solar radius
in antiphase with the solar cycle above $0.99\Rsun$,
but in phase between $0.97$ and $0.99\Rsun$.

The importance of using the $f$-mode frequencies for local
helioseismology has been recognized in a number of recent papers
\citep{HBBG08,DABCG11,FBCB12,FCB13}.
While such approaches should ultimately be used to determine the structure of
solar subsurface magnetic fields, we restrict ourselves here to the analysis
of oscillation frequencies as a function of horizontal wavenumber.
The purpose of the present paper is to use numerical simulations in piecewise
isothermal layers to study oscillation frequencies from random forcing
and to assess the effects of imposed magnetic fields on the frequencies.
In the present work we restrict ourselves to the case of uniform magnetic fields
and refer to the case of nonuniform fields in another paper \citep{SBR14},
where we study what is called a fanning out of the $f$-mode.

\section{Model and numerical setup}

\begin{figure}
\begin{center}
\includegraphics[width=\columnwidth]{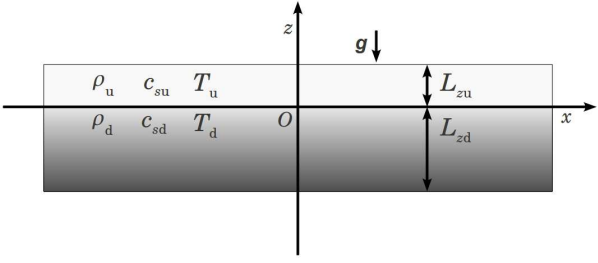}
\end{center}
\caption[]{
Geometry of the piecewise isothermal model.
The layer $z>0$ is hotter than the layer $z<0$.
}\label{box-2d}
\end{figure}

Let us consider a conducting fluid in a
two-dimensional Cartesian domain with $\ex$ and $\ez$ denoting the unit
vectors along the $x$ and $z$ directions, respectively.
Let gravity be acting along $-\ez$,
with constant acceleration $g>0$.
Thus we identify $x$ and $z$ as the horizontal and vertical directions, respectively.
Let the domain have a large horizontal extent $L_x$,
but a relatively small vertical extent $L_z\ll L_x$.
Due to gravity, the fluid is vertically stratified.
In addition it has an interface at $z=0$ with a
layer of thickness $\Lu$ above it, which is much
less dense than the layer of thickness $\Ld$ below\footnote{Subscripts
`u' and `d' indicate the value of a variable
in the `up' and `down' parts of the layer on both sides of the jump at $z=0$.}.
In the present work, we assume $\Lu=L_z/6$, so $\Ld=5L_z/6$.
At the interface, we assume a sharp jump in density $\rho$ with 
$\ru(0) \ll \rd(0)$
along with corresponding jumps in temperature and sound speed.
Such a setup may be thought of as 
an annular section with rectangular cross-section cut out from a star
with $z=0$ being its surface.
The thin layer on top represents then the
rarefied hot corona and the subdomain below $z=0$
stands for the more strongly stratified uppermost part of the convection zone,
which we sometimes refer to as the bulk.
A schematic diagram of this setup is shown in \Fig{box-2d}.

Assuming the fluid to obey the equation of state of an ideal gas,
the pressure is given by $p = (c_p-c_v) \rho T = \rho c_{\rm s}^2/\gamma$,
where $\gamma=c_p/c_v$ is the ratio of specific heats at constant pressure
and volume, respectively, and $\cs$ is the adiabatic sound speed.
Assuming further both subdomains to be {\it isothermal},
the scale heights of pressure and density are constant and equal
in each subdomain, $H_{p\rm d,u} = H_{\rho\rm d,u} = H_{\rm d,u}$.
Thus we have
\begin{align}
      \rho_{\rm d,u}(z) &=\rho_{\rm d,u}(0)\exp( -z/H_{\rm d,u} ),  \label{eq:backr} \\[-2mm]
\intertext{with}  \nonumber\\[-9mm] 
     H_{\rm d,u} &= (c_p-c_v)T_{\rm d,u}/g \,   \label{eq:backp}
\end{align}
The abrupt changes in the thermodynamic quantities at the interface
$z=0$ may be characterized by the ratio
\EQ
\Ratio=\ru(0)/\rd(0)=\csd^2/\csu^2
= H_{\rm d}/H_{\rm u}  = T_{\rm d}/T_{\rm u},
\EN
where pressure balance ($p_{\rm d} = p_{\rm u}$) at $z=0$ has been employed.
Different values of $\Ratio$ correspond to different factors by which
density, temperature and sound speed change abruptly at the interface.
For investigating a magnetic influence on the oscillation modes, we will
also consider an augmentation of the background state
by a uniform magnetic field $\BB_0=(B_{x0},0,B_{z0})$.
Note, that it does not affect the hydrostatic equilibrium.

\begin{figure}
\begin{center}
\includegraphics[width=\columnwidth]{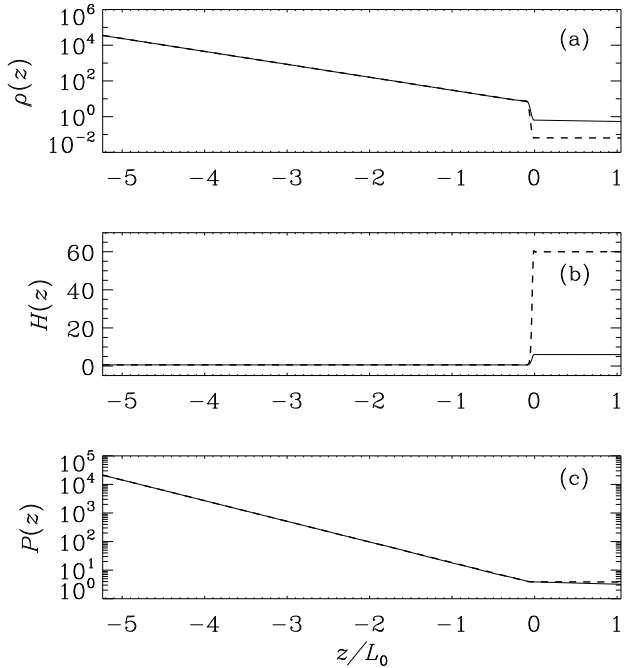}
\end{center}
\caption[]{
Density (a), pressure scale height (b), and pressure (c)
of the background state as functions of $z$
for $L_z/L_0=2\pi$. Solid: $\Ratio_0=0.1$,
dashed: $\Ratio_0=0.01$.
}\label{ther_var}
\end{figure}

In \Fig{ther_var} we show the variations of density,
pressure scale height and pressure as functions of $z$,
in a domain with 
$L_z/L_0=2\pi$ where $L_0= \gamma H_{\rm d}=\csd^2/g$.
The solid and dashed lines correspond to 
$\Ratio=0.1$ and $0.01$, respectively, which are
the two values employed in the present study.

Given that the oscillation modes are perturbations of the background state,
we solve in our DNS the time-dependent hydromagnetic equations, extended by
terms for both maintaining the background as well as exciting the oscillations.
Hence the basic equations adopt the form
\begin{align}
\frac{D \ln \rho}{Dt} &= -\bm\nabla\cdot\bm{u}, \\
\frac{D\bm{u}}{Dt} &= \ff+\bm{g} +\frac{1}{\rho}
\left(\bm{J}\times\bm{B}-\bm\nabla p
+\bm\nabla \cdot 2\nu\rho\bm{\mathsf{S}}\right),\\
T\frac{D s}{Dt} &= 2\nu \bm{\mathsf{S}}^2 + \frac{\mu_0\eta}{\rho}\JJ^2
- \gamma (c_p-c_v) \frac{T-T_{\rm d,u}}{\tau_{\rm c}}\, ,
\label{equ:ss} \\
\frac{\partial \bm A}{\partial t} &= {\bm u}\times{\bm B} - \eta \mu_0 {\bm J},
\end{align}
where $\bm{u}$ is the velocity, 
$D/Dt = \partial/\partial t + \bm{u} \cdot \bm\nabla$ is the
advective time derivative,
$\bm{f}$ is a forcing function specified below,
$\bm{g}=(0,0,-g)$ is the gravitational acceleration,
$\nu=\const$ is the kinematic viscosity, 
$\mathsf{S}_{ij}=\half(u_{i,j}+u_{j,i})
-\onethird \delta_{ij}\bm\nabla\cdot\bm{u}$
is the traceless rate of strain tensor, where commas denote
partial differentiation, ${\bm A}$ is the magnetic vector potential,
${\bm B} =\BB_0+\bm\nabla\times{\bm A}$ is the magnetic field,
${\bm J} =\mu_0^{-1}\bm\nabla\times{\bm B}$ is the current density,
$\eta=\const$ is the magnetic diffusivity, $\mu_0$ is the vacuum
permeability, and $T$ is the temperature.
For our purposes, viscous and Ohmic damping of the modes are 
parasitic effects, so we have included them only to ensure
numerical stability and chosen $\nu$ and $\eta$ as small as possible.
For the same reason, bulk viscosity was omitted.
The boundaries at $z=-L_{z\rm d},L_{z\rm u}$ are chosen to be impenetrable, stress-free,
and perfectly conducting,
while at the $x$ boundaries periodicity is enforced.

The last term in \Eq{equ:ss}, being of relaxation type, is
to guarantee that the temperature is on average constant in 
either subdomain and equal to $T_{\rm d}$ and $T_{\rm u}$, respectively.
For too high temperatures cooling is provided,
but heating for too low ones. 
Due to permanent viscous and Ohmic heating, however, cooling is necessary on average.
In test runs we found that the relaxation term can be omitted 
in the lower part of the domain
as long as the dissipation rate is sufficiently small.
Accordingly, the temperature in the lower subdomain is set by the initial condition
defined by the piecewise isothermal $z$ profiles with the desired value of $\Ratio$;    
see \Fig{ther_var}.
For numerical reasons, the jumps at $z=0$ were smoothed out over a few grid cells.
Note that without the relaxation term in $z<0$, there is a
slow drift in the $xt$ averages of density and temperature, reflected
by a deviation of the actual value of $\Ratio$,  
calculated with these averages,
from its initial value, which we now call $\Ratio_0$.
For the relaxation rate $\tau_{\rm c}^{-1}$ (in $z>0$) we choose
$0.5\,g/\csd$ throughout this paper.
On average, this corresponds to about $0.2\urms\kf$.

We provide a weak stochastic forcing $\ff$ in an
isotropic and homogeneous fashion, 
at a length scale, which is much shorter than the box dimensions,
for details see \cite{B01}.
The average forcing wavenumber $\kf$ defines the energy injection
scale $\lf=2\pi/\kf$ of the flow.
In the present study, we have used $\kf/k_1=20$, where
$k_1=2\pi/L_x$ is the lowest wavenumber in the domain.
Normally the forcing is specified to
act exclusively in the lower subdomain,
but we also compare with the case when it
is provided in the entire domain.
These two choices gave basically identical results,
most likely due to the weakness of the forcing and the fact that the
sound speed is high in the upper part and disturbances are quickly propagated.

We compute the root-mean-squared value of the turbulent motion from regions
below the interface ($z<0$), which we expect to be physically most relevant
for our purpose and denote it by $\ul$.
Let us define corresponding fluid Reynolds and Mach numbers of the flow as
\beq
\re \;=\; \frac{\ul}{\nu \kf},\qquad
\md \;=\; \frac{\ul}{\csd}.
\label{ReMa}
\eeq
For characterizing the forcing strength, it is useful
to employ the dimensionless quantity 
\beq
{\cal F} \;=\; f_0\,\rc \;=\; f_0\frac{\csd}{\nu\, \kf}\;,
\label{F}
\eeq
where $f_0$ is a dimensionless measure of the forcing amplitude 
\citep[non-dimensionalized by $\sqrt{\kf\csd^3/\delta t}$ with
the timestep of the numerical integration $\delta t$,][]{B01}.
$\rc$ may be thought of as the
fluid Reynolds number defined with respect to the sound speed $\csd$. 
The random forcing is expected to excite acoustic,
internal gravity, and surface waves,
referred to as $p\,$-, $g\,$-, and $f$-modes, respectively.
Although our primary goal is to study the properties
of the $f$-mode under a variety of physical conditions, we also 
turn to $p\,$-modes in some
detail, while $g\,$-modes are inspected only at a qualitative level.

It is customary to show the presence of these modes in a
$k_x$--$\omega$ diagram, to which we refer
in the following simply as $k\omega$ diagram.
It shows the amplitude of the Fourier transform of the vertical velocity $u_z$
as a function of $k_x$ and $\omega$.
Here, we take  $u_z$ from the interface at $z=0$, where the 
$f$-modes are expected to be most prominent.
By Fourier transforming $u_z(x,0, t)$ in $x$ and $t$, we obtain the
quantity $\hat{u}_z(k_x, \omega)$.
Results are presented in terms of the
dimensionless wavenumber $\widetilde{k}_x$ and 
angular frequency $\widetilde{\omega}$,
\EQ
\widetilde{k}_x = k_x L_0,\quad
\widetilde{\omega} = \frac{\omega}{\omega_0},
\quad \omega_0 = \frac{g}{\csd},
\label{DLko}
\EN
\noindent
where $\omega_0$, being twice the Lamb acoustic cutoff frequency 
of the bulk,
\EQ
\omega_c=g/2\csd=\csd/2L_0 ,  \label{Lambf}
\EN   
provides a natural time scale. 
As $\hat{u}_z(k_x, \omega)$ has the dimension of length squared,
we construct the dimensionless quantity 
\EQ
\widetilde{P}(\omega, k_x)\;=\;\frac{|\hat{u}_z|}{{\cal D}^2}
\;=\; \frac{|\hat{u}_z|}{L_0^2}\,\frac{\csd^2}{\ul^2},
\label{P}
\EN
where ${\cal D}=\ul/\omega_0$ characterizes the distance travelled 
with speed $\ul$ in an acoustic time $\omega_0^{-1}$.
%Note that ${\cal D}$ is smaller than $L_0$ by a factor $\Ma^2$.
%NS: modified
Note that ${\cal D}=\Ma L_0$ is smaller than $L_0$ for our values of $\Ma$.

\section{Non-magnetic case}
\label{nM}

We first study mode excitation in the absence of an imposed magnetic field by performing
simulations for three different extents ($L_x\times L_z$) of the domain,
while keeping $\Ratio_0=0.1$ and ${\cal F} = 0.05$ fixed; see \Tab{table-SnM}.
The different box sizes were chosen to compare properties such as frequency shifts
and line broadening of the $p\,$- and $f$-modes.

\begin{figure}%[t!]
\begin{center}
\includegraphics[width=\columnwidth]{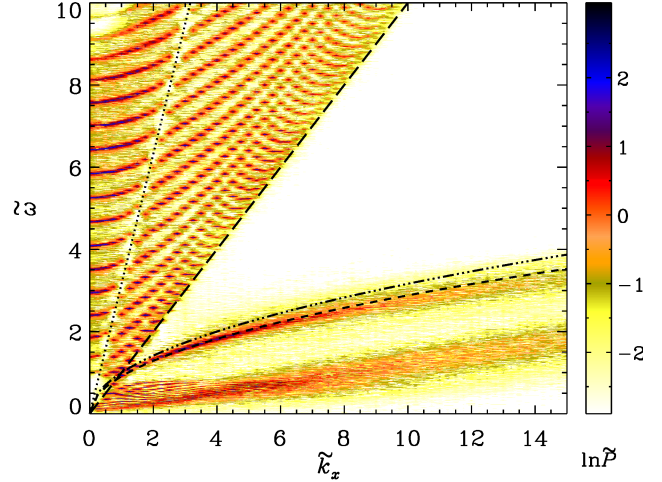}
\end{center}
\caption[]{$k\omega$ diagram for Run~A
($8\pi \times 2\pi$ domain, no magnetic field).
The dotted and long dashed
lines shows $\omega=\csu k_x$ and $\omega=\csd k_x$, respectively.
The dash-triple-dotted and dashed curves show $\omega_{\rm f0}$
and $\omega_{\rm f}$, respectively.}
\label{ko_GSnMa1}
\end{figure}

\begin{figure}%[t!]
\begin{center}
\includegraphics[width=\columnwidth]{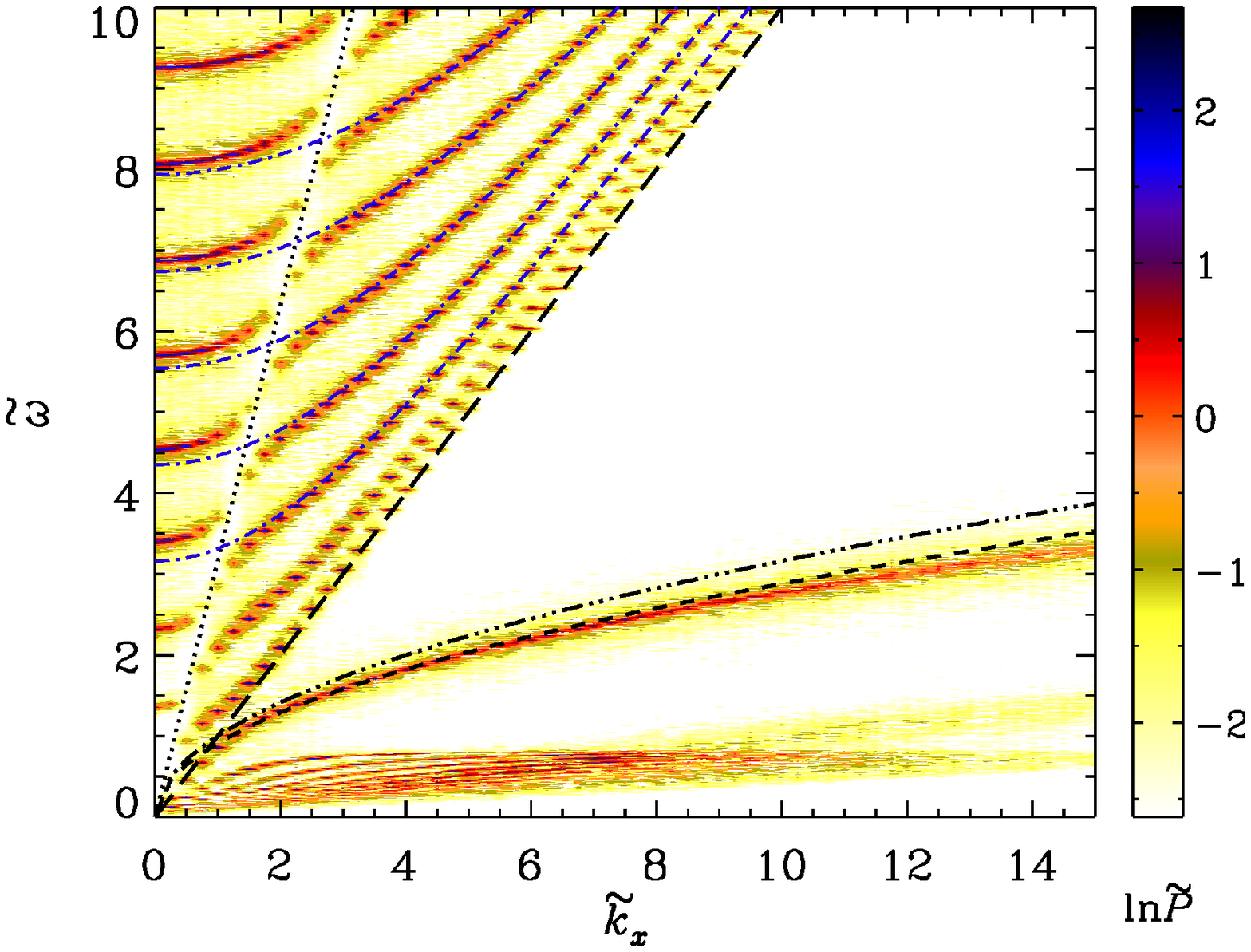}
\end{center}
\caption[]{Same as \Fig{ko_GSnMa1} but for Run~B
($8\pi \times \pi$, no magnetic field).
Dash-dotted (blue): some of the estimated $p\,$-modes according to
\Eqs{disp}{Quarter-Wave}.
}
\label{ko_GSnMa1_8pixpi}
\end{figure}

The $k\omega$ diagrams for Runs~A ($L_z=2\pi$) and B ($L_z=\pi$)
are shown in \Figs{ko_GSnMa1}{ko_GSnMa1_8pixpi},
where modes of all types ($p$, $g$ and $f$) clearly appear.
The multiple curves to the left of the long-dashed line belong to
the $p\,$-modes, the curve just below the short-dashed line corresponds
to the $f$-modes and the curves further below indicate $g\,$-modes.
Especially in the lower left corner of \Fig{ko_GSnMa1_8pixpi}
one can distinguish several different branches.

\begin{table}\caption{
Summary of simulations without magnetic field,
$\Ratio_0=0.1$ and ${\cal F}=0.05$.
}\centering
\label{table-SnM}
\resizebox{\columnwidth}{!}{%
\begin{tabular}{c l r c c c c }
\hline\hline
Run & Domain & Grid\phantom{600} & $\Ratio$ & $\re$ & $\md$\\ [-.2ex]
\hline
A & $8\pi \times 2\pi$ & $1024\times 600$ & 0.093 & 1.94  & 0.004 & (\Fig{ko_GSnMa1})\\
B & $8\pi \times \pi$ & $1024\times 300$ & 0.092 & 0.95  & 0.002 & (\Fig{ko_GSnMa1_8pixpi})\\
C & $4\pi \times \pi$ & $512\times 300$ & 0.091 & 0.93  & 0.002 \\
\hline
\end{tabular}%
}
\end{table}

\subsection{$p\,$-modes}

The $p\,$-modes, also known as pressure modes, are acoustic waves
that are trapped in a resonant cavity.
In a stratified isothermal medium (without interface),
their dispersion relation is in two dimensions 
approximately given by
\EQ
\omega^2 \approx \omega_c^2+\cs^2\left(k_x^2+k_z^2\right),  \label{disp}
\EN
where $k_x$ and $k_z$ adopt discrete values depending on
the extent of the cavity and $\omega_c=g/2\cs$ is again
the Lamb acoustic cutoff frequency \eqref{Lambf}.
In \Eq{disp} the contribution
$N^2/\omega^2$ with the Brunt-V\"ais\"al\"a frequency $N=(\gamma-1)^{1/2} g/\cs$
\citep{SL74} has been ignored, 
as typically $\omega\gg N$ for acoustic modes. %MR: but maybe not for p_0?
Impenetrable $z$ boundaries let the waves be standing
in the $z$ direction, whereas periodic $x$ boundaries
allow them to travel in the $x$ direction.
For the vertical wavenumber $k_z$, the discrete
values $n\pi/L_z$ with integer $n$ are possible
and thus from \Eq{disp} a set of
eigenvalue curves $\omega_n(k_x)$ can be formed.
In a $k\omega$ diagram they are bound
from below by the asymptotic line $\omega=\cs k_x$.  

Turning to our two-layer setup and inspecting
the $k\omega$ diagrams in \Figs{ko_GSnMa1}{ko_GSnMa1_8pixpi},
we note that the asymptotic line is now found to be
$\omega= \csd k_x$, just as the cavity were
given by the lower subdomain.
As a major difference from the picture
expected for a single layer, gaps coinciding
with apparent discontinuities in the 
$\omega_n(k_x)$ curves occur. They seem to line up along 
\EQ
\omega= \csu k_x,   \label{eq:sepline}
\EN
which would be the asymptotic line if the cavity
were given by the upper subdomain alone.
The higher complexity of the $k\omega$ diagram in comparison with
the single layer is due to the existence of three different families of
$p$-modes instead of only one:
while in a single isothermal layer all standing waves have the $z$
dependence $\exp(-\kappa z) \sin( k_z z + \phi)$, the two-layer setup
allows additionally waves which are purely exponential (or evanescent)
in one of the subdomains and (damped) harmonic in the other.
Those which are evanescent in the bulk are in \cite{HZ94}
identified with the line \eqref{eq:sepline} and called ``$a$-modes'',
although the actual mode frequencies follow a pattern that is referred
to as ``avoided crossings'' with the $p$- and $f$-modes.
In the following we refer to this line as separatrix.
The suppression of the mode amplitude around the separatrix is, however,
not explained by ideal linear theory.

\begin{figure}%[t!]
\begin{center}
\includegraphics[width=\columnwidth]{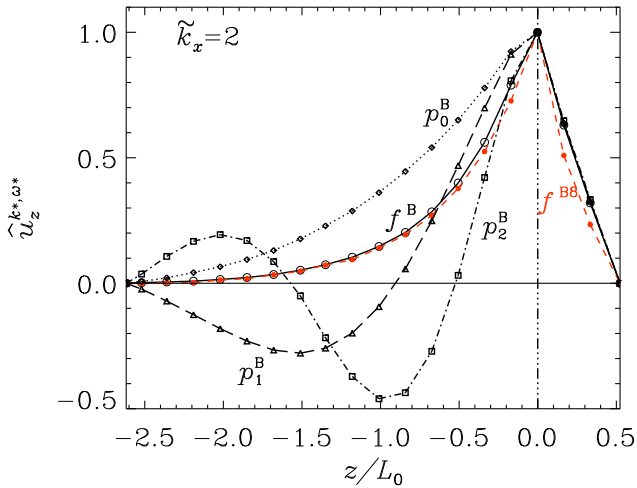}
\end{center}
\caption[]{
Eigenfunctions of the $f$- and the first three $p$-modes
($p_0$, $p_1$ and $p_2$) at $\widetilde{k}_x=2$.
Black curves correspond to Run~B,
whereas the red curve (with filled circles) shows the $f$-mode
corresponding to Run~B8, which has horizontal magnetic field.
The vertical (dash-dots) line shows the location of the interface.
}
\label{EFs_p-f}
\end{figure}

To determine the shape of the $f$- and $p$-mode eigenfunctions,
we derived them from the $z$ dependent spectrum of $u_z$ by selecting
$\widetilde{k}_x=2$ and $\widetilde{\omega}=1.31,\,2.09,\, 2.77,\, 3.67$
corresponding to the $f$-mode and the first three $p$-modes, $p_{0,1,2}$,
respectively.
The result is shown in \Fig{EFs_p-f}, corresponding to
Runs~B and B8.
For $p_0$ the eigenfunction is approximately a (damped) quarter wave
with a node at the bottom and a maximum at the interface.
Accordingly, we make the following tentative ansatz for
the vertical wavenumbers (analogous to those of organ pipes):
\EQ
k_z\;=\;\frac{\pi (n+1/2)}{L_{z\rm d}}\;,\quad
n=0,1,2,\ldots\,,
\label{Quarter-Wave}
\EN
where $n$ is the number of nodes in the $z$ direction.
It is interesting to recall here Duvall's law, which was applied to
the solar observations \citep{D82}:
\EQ
k_z\;=\;\frac{\pi (n_D+\alpha)}{L_{z\rm d}}\;,\quad
n_D=1,2,3,\ldots\,.
\label{Duvall}
\EN
The fundamental mode, with radial node number $n_D=0$, was excluded
from this formula, and the best fit value for $\alpha$ was found to be
$\approx3/2$, which is intended to
account for the fact that the interface is ``soft''
instead of rigid.
In general, $\alpha$ has to be considered frequency dependent
\cite[see][]{G87,C03}.
We also note that the possibility of misidentification of the radial node
number ($n_D$) was explicitly discussed in \cite{D82}.

Given that our model setup is quite different from the real Sun,
we should perhaps not seek direct comparison between
\Eqs{Quarter-Wave}{Duvall}, although it is noteworthy to 
mention that, as shown in \Fig{EFs_p-f}, 
both the $f$- and the $p_0$- mode, corresponding to $n=0$
in \Eq{Quarter-Wave}, do not have any node within the bulk.
Hence one could ask whether a qualitative distinction between those
modes is tenable.
This might help resolving the observational issue of proper identification
of the radial node number, when the eigenfunctions of various trapped
modes are not readily accessible. Also, note that \Eqs{Quarter-Wave}{Duvall}
are equivalent if $\alpha=1/2$ and $n_D$ is allowed to include
the value $0$, too.

\begin{figure}
\begin{center}
\includegraphics[width=\columnwidth]{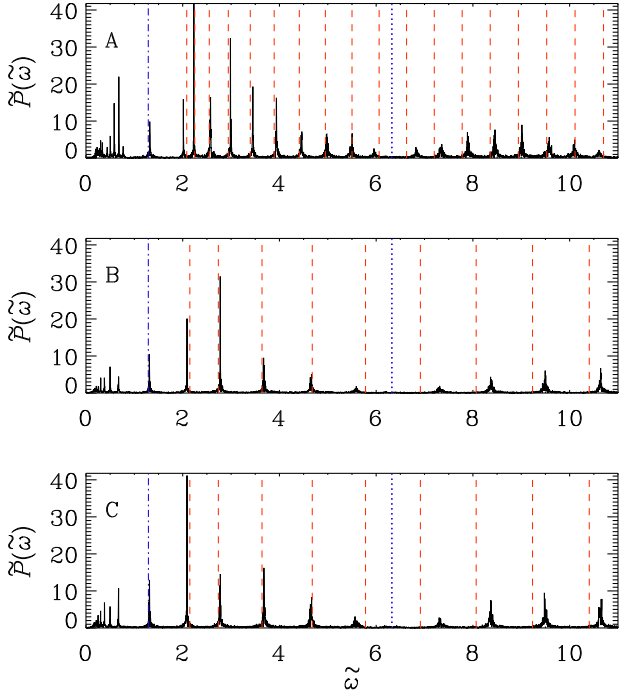}
\end{center}
\caption[]{$\widetilde{P}(\widetilde{\omega}; \widetilde{k}_x=2)$
as a function of $\widetilde{\omega}$ in the absence of
a magnetic field for Runs~A, B and C (top to bottom);
see \Tab{table-SnM}.
Dash-dotted (blue) and dashed (red):
theoretical locations of the $f$- and $p\,$-modes
according to Eqs~(\ref{DR-NM}), (\ref{disp}) and (\ref{Quarter-Wave}),
respectively.
(Blue) dotted lines: position of the separatrix \eqref{eq:sepline}
shown dotted in the $k \omega$ diagrams (see e.g., \Fig{ko_GSnMa1}).}
\label{fp-pow_kx2-nM}
\end{figure}

In \Figs{fp-pow_kx2-nM}{fp-pow_kx4-nM} we plot
the dimensionless spectral mode amplitude $\widetilde{P}$
as a function of $\widetilde{\omega}$ for $\widetilde{k}_x=2$
and $4$, respectively, for Runs~A, B and C; see also \Tab{table-SnM}.
The dash-dotted (blue) and dashed (red) lines mark the 
locations of the $f$- and $p\,$-modes as expected from
\Eq{DR-NM} and \Eq{Quarter-Wave}, respectively.
The group of peaks to the left of the $f$-mode indicate $g\,$-modes.
The (blue) dotted lines mark 
the $\omega=\csu k_x$ line, which is shown by the dotted line
in \Figs{ko_GSnMa1}{ko_GSnMa1_8pixpi}.

\begin{figure}
\begin{center}
\includegraphics[width=\columnwidth]{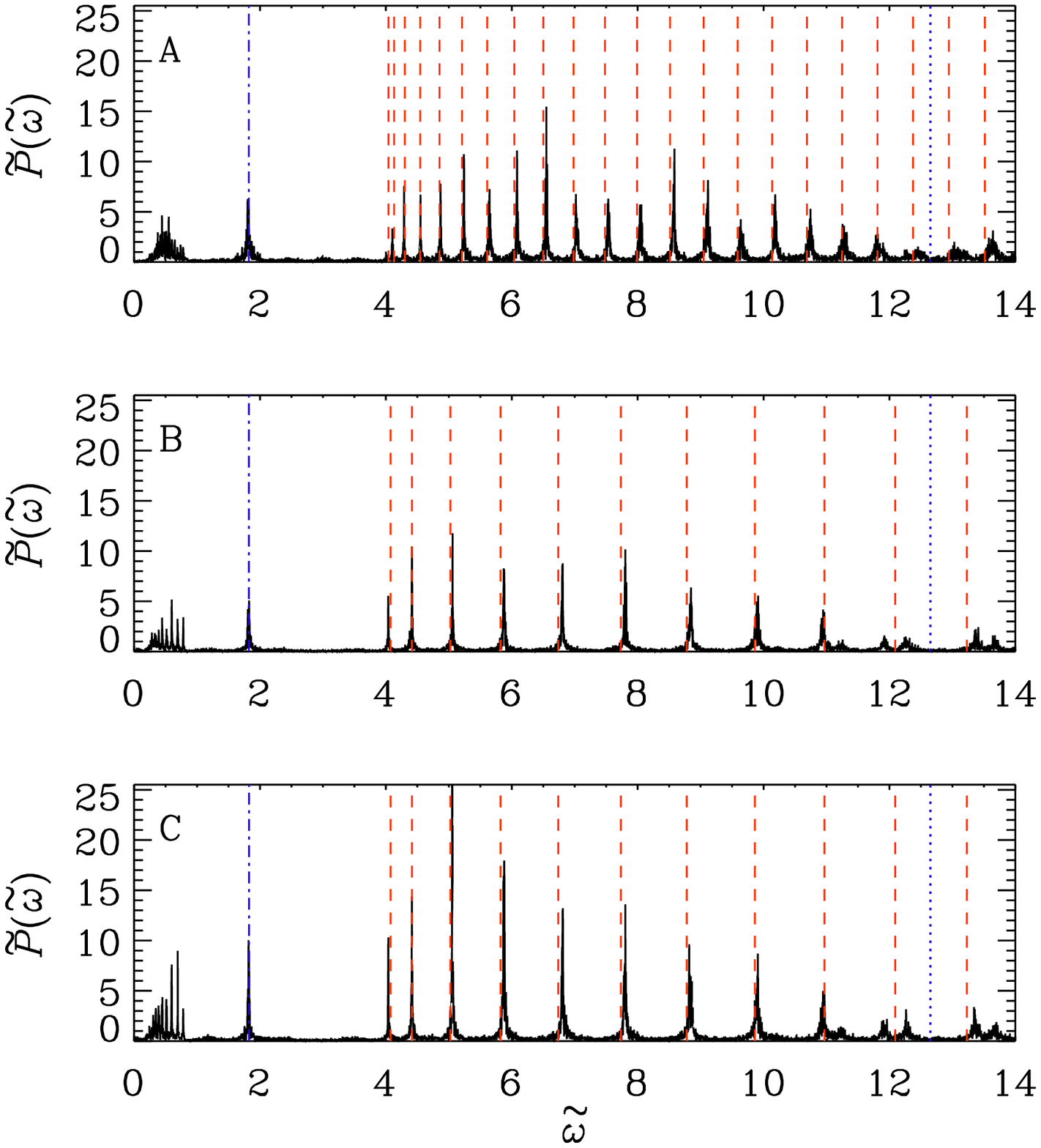}
\end{center}
\caption[]{Same as \Fig{fp-pow_kx2-nM}, but for $\widetilde{k}_x=4$.}
\label{fp-pow_kx4-nM}
\end{figure}

We find that various peaks of the $p\,$-mode appear at the 
locations predicted by \Eq{Quarter-Wave}, although
there are some slight frequency shifts, as may be seen from
\Figs{fp-pow_kx2-nM}{fp-pow_kx4-nM}.
We also note that the frequency shift changes sign across
the $\omega=\csu k_x$ line, shown by
blue dotted lines in \Figs{fp-pow_kx2-nM}{fp-pow_kx4-nM}.
%AB: moved the following from below to here (as instructed in proofs corrections)
Comparing the different panels in each of 
\Figs{fp-pow_kx2-nM}{fp-pow_kx4-nM}, we first note that the number of
$p\,$-mode peaks are two times smaller in 
Runs~B and C compared to Run~A.
This is expected as the vertical extent ($L_z$) in Run~A is twice as large
as in Runs~B and C.
From \Fig{fp-pow_kx2-nM}, corresponding to $\widetilde{k}_x=2$,
we see that the dimensionless mode amplitudes $\widetilde{P}(\omega)$ of the
corresponding peaks of the $p\,$-mode are in all three cases comparable,
even though the domain sizes are different.
The same holds true for \Fig{fp-pow_kx4-nM} corresponding
to $\widetilde{k}_x=4$, although the mode amplitudes have generally
decreased for the $\widetilde{k}_x=4$ case.
We find that the Mach number in the lower subdomain decreases by the
same factor as we decrease the vertical extent of the cavity in spite
of the same strength of forcing in all cases (see \Tab{table-SnM}
and note that ${\cal F}=0.5$ for all three cases).
This might be due to the fact that a larger number of $p\,$-modes
are present in a larger cavity (in our case, twice as many), 
thus contributing
to the random motion, which increases the value of $\md$.

Frequency shifts of the $p\,$-modes due to the presence of a hot layer
above the cavity have been calculated in \cite{HZ94,DSA00}.
The first of these papers deals with $p$-modes in an infinitely
extended volume filled with a gas obeying an adiabatic equation of state.
Temperature and hence sound speed are considered height
dependent with a minimum close to the solar surface. Eigensolutions are
obtained by requiring regularity at infinity.
The second work employs a piecewise isothermal three-layer model
in which between two layers, allowing sound waves to propagate,
a further one is embedded where the waves are evanescent. This
setup resembles the conditions in the neighbourhood of the
temperature minimum in the lower chromosphere. It is
demonstrated that, if the cool layer is sufficiently shallow,
$p$-modes can ``tunnel'' through it from one propagation layer
to the other. The results of both papers cannot easily be applied
to our two-layer model.

\subsection{$f$-mode}
\label{fmode_nonmag}

The classical $f$-mode, also known as the {\it fundamental mode}, is a surface
wave which exists due to a discontinuity in the density.
In the absence of a magnetic field, the dispersion relation for the
$f$-mode is given by \citep[see, e.g.,][]{G87,CR89,ER90}
\beq
\omega_{\rm f}^2 = g k_x \, \frac{1-\Ratio}{1+\Ratio}, \quad
\omega_{\rm f0}^2 = g k_x,
\label{DR-NM}
\eeq
where $\omega_{\rm f0}$ is the frequency in the limit when $\ru=0$, 
that is, $\Ratio=0$.
The dashed and dash-triple-dotted curves in \Figs{ko_GSnMa1}{ko_GSnMa1_8pixpi} show
$\omega_{\rm f}$ and $\omega_{\rm f0}$, respectively.
For linear perturbations,
the dispersion relation as given by
\Eq{DR-NM} is independent of the background
stratification and the thermodynamic properties of the fluid.
Consequently the $f$-mode, in contrast to the $p\,$-modes,
was traditionally expected to be less of diagnostic value, and received less attention.
However, observations \citep{FSTT92,DKM98}
revealed significant deviations of the frequencies of the solar $f$-modes
from the simple relation given in \Eq{DR-NM}, so their diagnostic importance
grew significantly \citep{RG94,RC95,GAC95,MMR99,M00a,M00b}.
Attempts were made to explain the frequency shifts of the high spherical 
harmonic degree solar $f$-modes by considering them as interfacial waves,
which propagate at the chromosphere-corona transition \citep{RG94,RC95}.
The other hypothesis mentioned earlier invokes
frequency shifts and line broadening of the $f$-mode
due to the turbulent motions \citep{MR93a,MR93b,MMR99,M00a,M00b}.
It is at present unclear which of these explanations 
is more relevant for the Sun.

In our simulations, the $f$-mode frequencies lie significantly below
$\omega_{\rm f0}$ and are much closer to $\omega_{\rm f}$ for small $k_x$
values (see \Figs{ko_GSnMa1}{ko_GSnMa1_8pixpi}).
However, for large values of $k_x$ the line center falls progressively
below the theoretically expected $\omega_{\rm f}$ curve.
We also notice a line broadening for large values of $k_x$.
It is discussed in earlier works \citep{MMR99,M00a,M00b} that,
for large horizontal wavenumbers, both frequency shift and line broadening 
may be caused by incoherent background motions that we simply refer to as ``turbulence''.
However, it may be more appropriate to characterize 
shift and broadening
as finite Mach number effects.

\begin{figure}
\begin{center}
\includegraphics[width=\columnwidth]{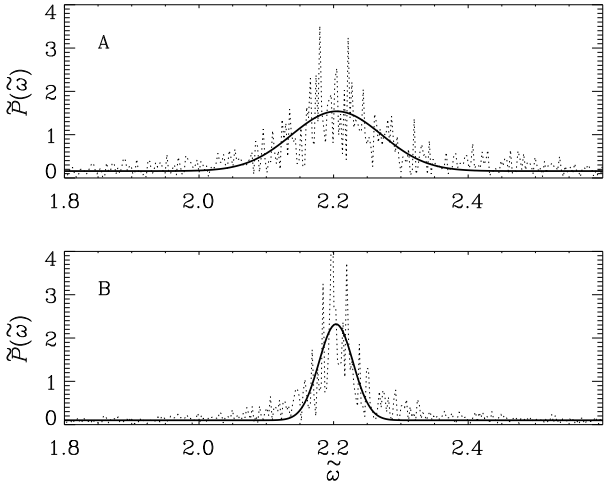}
\end{center}
\caption[]{Line profiles of the $f$-mode at $\widetilde{k}_x=6$
in the \emph{absence} of a magnetic field
for Runs~A and B.
Dotted: data from DNS, solid: Gaussian fit.
}
\label{fmline-nM}
\end{figure}

In order to quantify our results, we focus on the line profiles of the $f$-mode
under various physical conditions. 
To that end, we fit the quantity $\widetilde{P}(\omega,k_x)$ for a fixed value
of $k_x$ to a Gaussian by a robust nonlinear least squares
method using publicly available standard procedures \citep{M09}.
The fit parameters include the central frequency $\ofn$,
the line width, the peak value and the vertical shift.
In \Fig{fmline-nM}, dotted lines  indicate $\widetilde{P}(\omega,k_x)$
at $\widetilde{k}_x=6$ for Runs~A and B, 
while the Gaussian fit is given by bold lines.
The vertical shift characterizes the `turbulence continuum',
whereas the Gaussian profile represents the mode proper.

Let us denote the numerical estimate of the line center from the fit
by $\ofn$ and characterize the {\it relative frequency shift} by
\beq
\frac{\delta \omega_{\rm f}^2}{\omega_{\rm f}^2} \equiv
\frac{\oofn-\omega_{\rm f}^2}{\omega_{\rm f}^2}
\label{do2f}
\eeq
as a measure of the departure 
of the detected $f$-mode frequency from its theoretical value \eqref{DR-NM}.
For characterizing the amplitude of an $f$-mode
we define the {\it normalized mode mass} as the area under the
Gaussian fit after subtracting the continuum:
\beq
\mu_{\rm f} \;=\; \frac{1}{\nut}
\int \Delta|\hat{u}_z |\, \mathrm{d}\omega, 
\label{muf}
\eeq
where $\Delta|\hat{u}_z|$ denotes the excess over the `continuum'
and $\nut=\ul/3 \kf$ is an estimate for the turbulent viscosity,
which has been used on purely dimensional grounds.
The dimensionless measure of the {\it linewidth} of an $f$-mode at a
central frequency $\ofn$ is defined by
\beq
\Gamma_{\rm f} \;=\; \frac{\Delta \omega_{\rm FWHM}}{\ofn},
\label{lwf}
\eeq
where $\omega_{\rm FWHM}$ is the full width at half maximum of the
line profile.
Frequency shift, mode mass and linewidth
will now be employed to analyse DNS under
a variety of physical conditions.

\begin{figure}
\begin{center}
\includegraphics[width=\columnwidth]{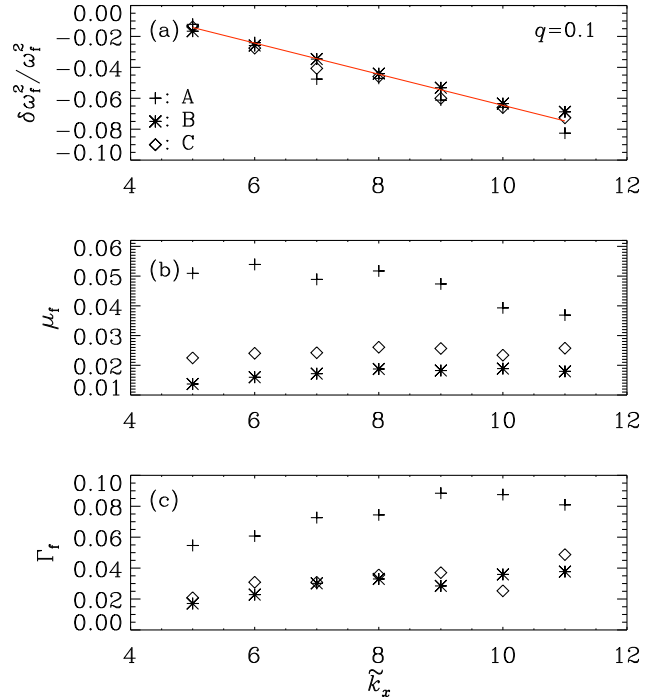}
\end{center}
\caption[]{
Relative frequency shift (a), mode mass (b), and linewidth (c) of
the $f$-mode as a function of $\widetilde{k}_x$ in the absence of a
magnetic field.
Symbols `$+$', `$\ast$', and `$\diamond$':
Runs~A, B and C, respectively (see \Tab{table-SnM})
and the red line represents an approximate fit to the data.
}\label{fmode-nM}
\end{figure}

Noting the fact that the Mach number is larger in a deeper domain
(see \Tab{table-SnM}), we find that
(i) the peak value of $\widetilde{P}(\widetilde{\omega})$ decreases
with increasing $k_x$ and is larger for a shallower domain at
any value of $k_x$ than for a deeper one,
(ii) the linewidth of the $f$-mode
increases with $k_x$ and is smaller for a shallower domain.

In \Fig{fmode-nM} we show the dependencies of
$\delta\omega_{\rm f}^2/\omega_{\rm f}^2$, 
$\mu_{\rm f}$, and $\Gamma_{\rm f}$
on $k_x$ for the Runs~A, B and C,
which have different domain sizes and Mach numbers (see \Tab{table-SnM}).
Remarkably, the values of $\delta\omega_{\rm f}^2/\omega_{\rm f}^2$
for Runs~A, B and C almost coincide in \Fig{fmode-nM}(a)
and show a linearly decreasing trend with
$k_x$, indicated by the red line with slope $\sim-0.01$.
However, such $k_x$-dependent frequency shifts of the
$f$-mode are expected both from the influence of
turbulent motions 
\citep[][where shifts are linear in $k$]{MMR99,M00a,M00b,MKE08},
as well as for interfacial waves propagating in the transition region
between chromosphere and corona \citep{RG94,RC95}.
By comparison, purely viscous effects proportional to $\nu k^2$
would yield a relative frequency shift
$\delta\omega^2/\omega^2\approx 2\nu k_x^{3/2}/\sqrt{g}$,
which is around $10^{-3}$ for $\widetilde{k}_x=10$.
Although this correction would 
be linear in $k_x$, just like in our simulations,
it is clearly negligible.
Note also that the mode mass decreases only slightly with
$k_x$ for the deeper domain, while for the shallower one,
however, it stays nearly constant.
The linewidth increases in all cases somewhat with $k_x$.
Also, both $\mu_{\rm f}$ and $\Gamma_{\rm f}$ are considerably smaller
in the shallower domains of Runs~B and C, compared to the deeper one of Run~A,
but it is worthwhile to note that
the corresponding Mach numbers are also smaller, as will be discussed in \Sec{fm_mode}.
We return to this in the next section, where we show that in our simulations magnetic fields
also affect the Mach number and thus the mode mass.

\subsection{$g\,$-modes}

When inspecting \Figs{ko_GSnMa1}{ko_GSnMa1_8pixpi}, 
we note that the $g\,$-modes appear sharper in the latter case
(Run~B, shallower domain) compared to the former (Run~A, deeper domain).
Individual $g\,$-modes can be identified for small values of $\widetilde{k}_x$
up to $\tilde{k}_x=4$ and 6, respectively.
The envelope surrounding the g modes shows saturation for $\widetilde{k}_x \lesssim 7$ and 4, respectively.
For larger values of $\widetilde{k}_x$, the envelope
% surrounding the $g\,$-modes 
seems to continue with a mild increase in frequency.
Although this effect is more clearly seen for the deeper
domain (\Fig{ko_GSnMa1}), in both cases the center line is
nevertheless compatible with a straight line going through zero
with almost the same slope and almost the same intersection point,
$(\widetilde{k}_x,\widetilde{\omega})\approx(7,0.8)$, with the saturated
part of the
envelope.
% of the $g$-modes corresponding to the lower subdomain.
%The saturation of the $g$-mode frequencies with respect to $k_x$
%is expected
The latter can be explained
 from the analysis of a single isothermal layer,
 applied to the lower subdomain.
A further linear increase of the $g$-mode frequencies
is possible due to the existence of $g$-modes in the upper subdomain
of a two-layer setup; see \cite{HZ94}.
These are also expected to saturate at some constant,
but for much higher values of $k_x$.
\cite{WS07}, while studying the effect of a hot layer above the
photosphere, find that there is a continuous spectrum below the
$f$-mode, with frequencies linearly increasing with $k_x$ for
intermediate values of $\widetilde{k}_x$.

\section{Horizontal magnetic field}
\label{HM}

\begin{figure}
\begin{center}
\includegraphics[width=\columnwidth]{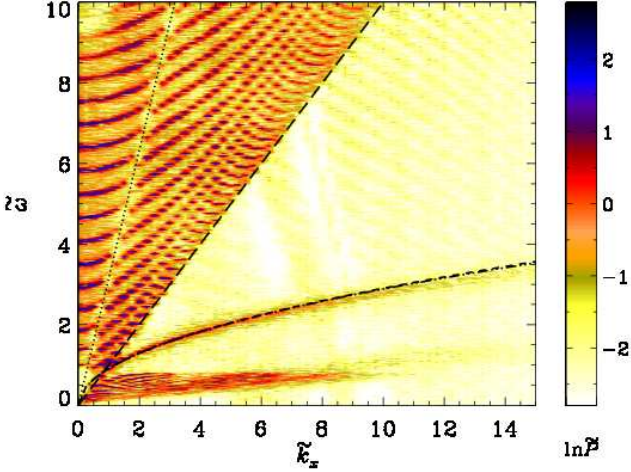}
\end{center}
\caption[]{$k\omega$ diagram for Run~A2
with horizontal magnetic field.
The dotted and long dashed lines show $\omega=\csu k_x$ and
$\omega=\csd k_x$, respectively.
The dash-triple-dotted and dashed curves
(nearly on top of each other as
$B_{x0}$ is small; but cf. \Fig{ko_TSMb2})
show $\omega_{\rm f}$ and $\omega_{\rm fm}$,
respectively; see \Tab{table-SMH}.
}
\label{ko_GSMa1-2}
\end{figure}

\begin{figure}
\begin{center}
\includegraphics[width=\columnwidth]{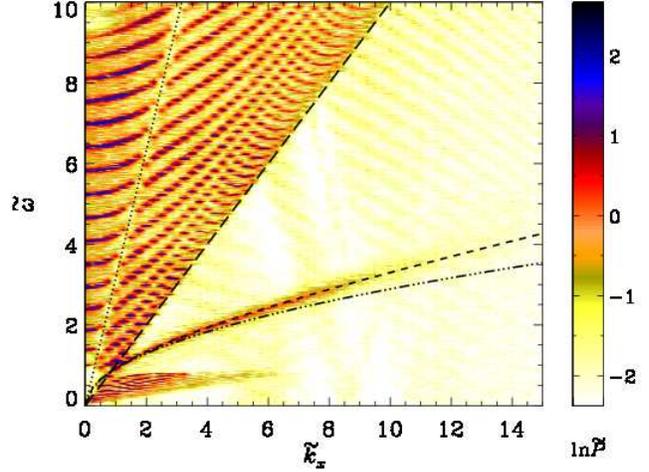}
\end{center}
\caption[]{Same as \Fig{ko_GSMa1-2},
but for Run~A5 with a 5 times stronger magnetic field.}
\label{ko_TSMb2}
\end{figure}

We impose a uniform horizontal magnetic field $(B_{x0},0,0)$ in the entire
domain and study its effect on the different modes
for varying domain sizes, density jumps, and field strengths $B_{x0}$,
with a focus on its effect on
the $f$-mode.
Details of the simulations
discussed in this section are summarized in \Tab{table-SMH},
where we have employed Alfv\'en velocities
which are different in the bulk and the corona according to
\beq
v_{{\rm A}x\,{\rm d,u}} \,=\, B_{x0}/\sqrt{\mu_0\rho_{\rm d,u}(0)}.
\label{vAx-ud}
\eeq
On the other hand, the ratio $v_{{\rm A}x}/\cs$ is approximately the same
above and below the interface.
So, for definitiveness, we denote by $\vA/\cs$ in the following
the average of both values.

The $k\omega$ diagrams for some of our runs are shown in
\Figss{ko_GSMa1-2}{ko_GSMa89}.
The $p\,$-, $g\,$- and $f$-modes appear clearly in all the 
diagrams
and are seen to be affected by the
magnetic field.
As before, the dotted and long-dashed lines show $\omega=\csu k_x$ and
$\omega=\csd k_x$, respectively.

\begin{table*}\caption{
Summary of simulations with horizontal magnetic field. 
Runs~A1--A6h: $L_z/L_0=2\pi$; Runs B8 and B8h: $L_z/L_0=\pi$.
}\centering
\label{table-SMH}
\begin{tabular}{l c c c c c c c c c c c c}
\hline\hline
Run & Grid & $\Ratio_0$ & $\Ratio$ & $\vaxd/\csd$ & $\vaxu/\csu$ & $\md$ & $\re$ & ${\cal F}$\\ [-0.2ex]
\hline
A1 & $1024\times 600$ & 0.1 & 0.097 & 0.004 & 0.004 & 0.0030 & 0.71 & 0.025\\
A2 & $1024\times 600$ & 0.1 & 0.091 & 0.025 & 0.025 & 0.0032 & 0.61 & 0.020 & (\Fig{ko_GSMa1-2})\\
A3 & $1024\times 600$ & 0.1 & 0.095 & 0.042 & 0.041 & 0.0027 & 0.52 & 0.020\\
A3'& $1024\times 600$ & 0.1 & 0.093 & 0.042 & 0.041 & 0.0028 & 0.54 & 0.020\\
A4 & $1024\times 600$ & 0.1 & 0.095 & 0.106 & 0.102 & 0.0029 & 0.68 & 0.025\\
A5 & $1024\times 512$ & 0.1 & 0.089 & 0.124 & 0.123 & 0.0032 & 0.76 & 0.025 & (\Fig{ko_TSMb2}) \\
\hline
A3h& $1024\times 600$ & 0.01 & 0.0110 & 0.042 & 0.039 & 0.0014 & 0.27 & 0.020 & (\Fig{ko_GSMa4}) \\
A5h& $1024\times 512$ & 0.01 & 0.0091 & 0.119 & 0.116 & 0.0021 & 0.50 & 0.025\\
A6h& $1024\times 512$ & 0.01 & 0.0096 & 0.162 & 0.156 & 0.0015 & 0.34 & 0.020 & (\Fig{ko_TSMb3}) \\
\hline\hline
B8 & $1024\times 300$ & 0.1 & 0.099 & 0.296 & 0.283 & 0.0016 & 0.78 & 0.050 & (\Fig{ko_GSMa89}a) \\
B8h& $1024\times 300$ & 0.01& 0.0098& 0.290 & 0.273 & 0.0015 & 0.71 & 0.050 & (\Fig{ko_GSMa89}b) \\
\hline
\end{tabular}%
%}
\end{table*}

%% \Ratio=0.01
\begin{figure}%[t!]
\begin{center}
\includegraphics[width=\columnwidth]{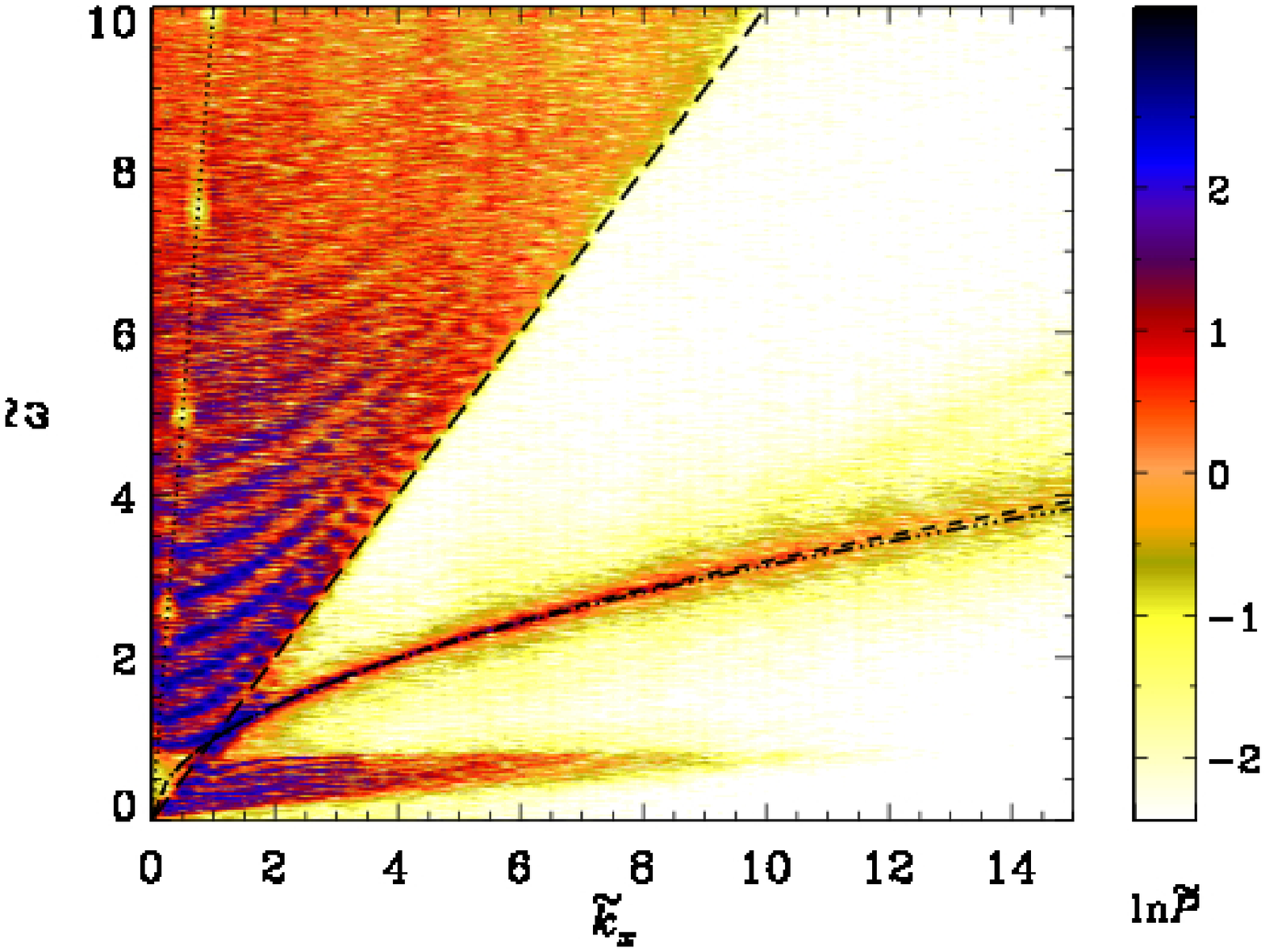}
\end{center}
\caption[]{Same as \Fig{ko_GSMa1-2},
but for Run~A3h with hotter corona.}
\label{ko_GSMa4}
\end{figure}

\begin{figure}%[t!]
\begin{center}
\includegraphics[width=\columnwidth]{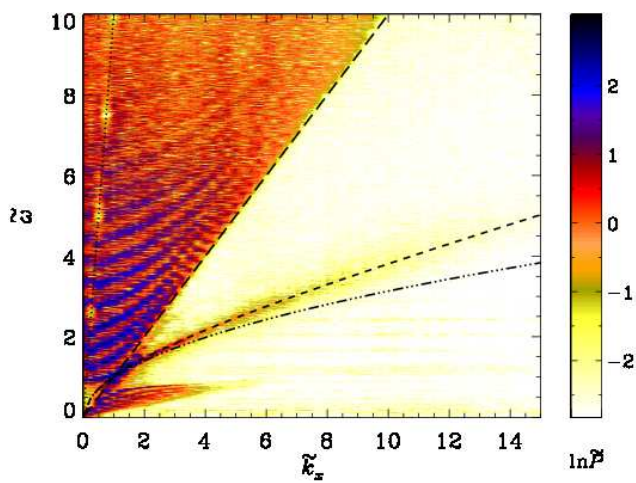}
\end{center}
\caption[]{Same as Fig.~(\ref{ko_GSMa4}),
but for Run~A6h with a stronger magnetic field.}
\label{ko_TSMb3}
\end{figure}

\begin{figure}%[t!]
\begin{center}
\includegraphics[width=\columnwidth]{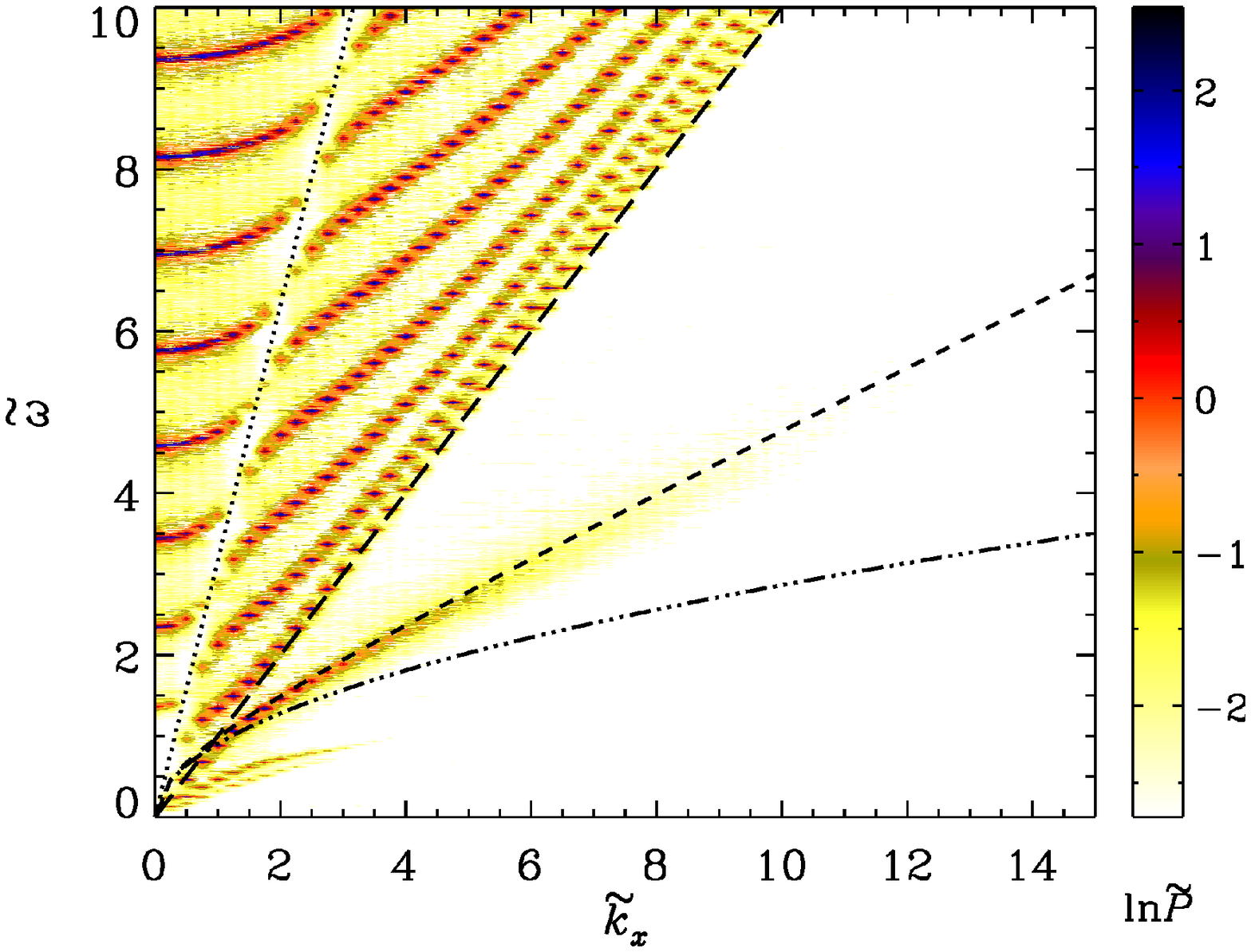}
\includegraphics[width=\columnwidth]{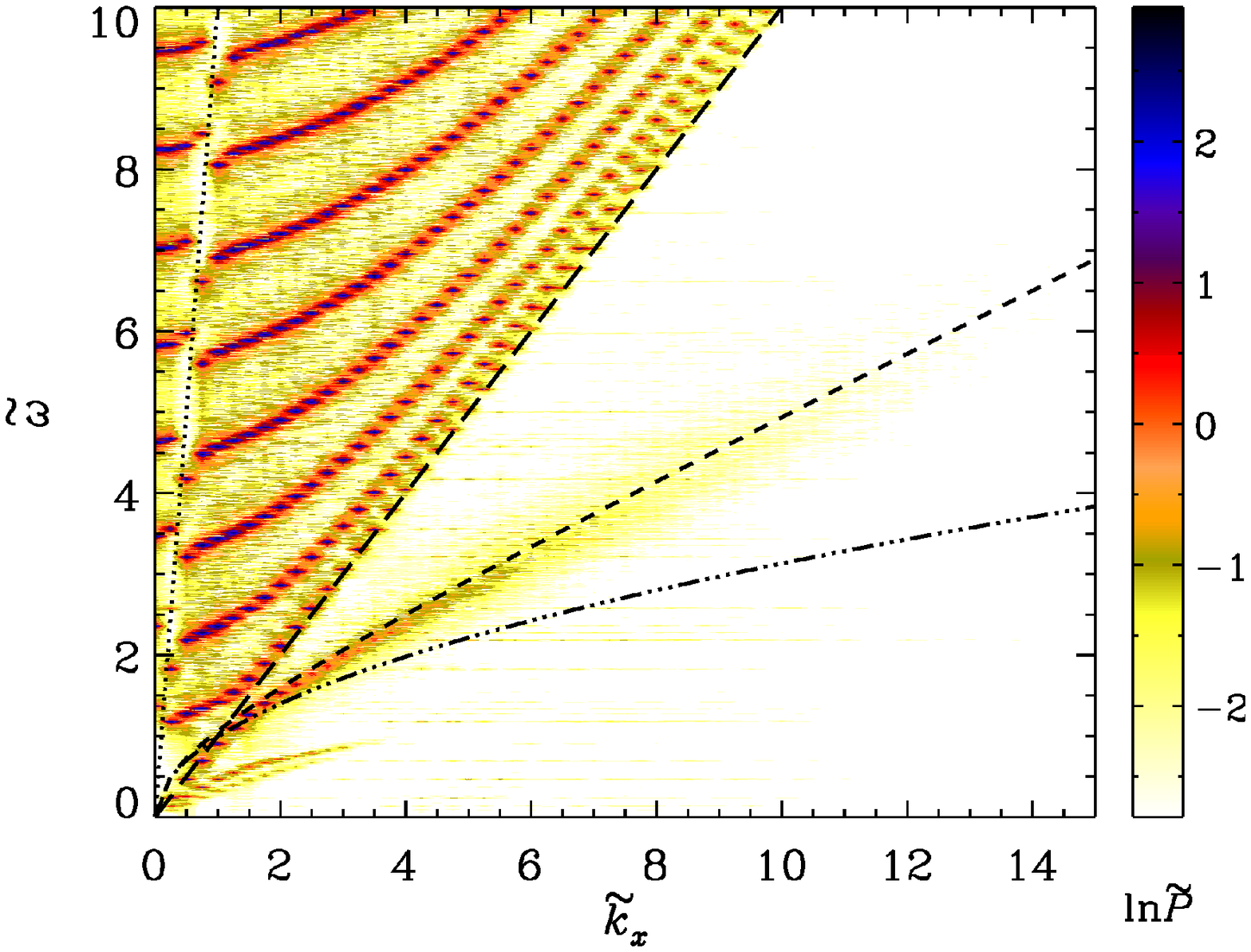}
\end{center}
\caption[]{$k\omega$ diagrams for a domain $8\pi\times \pi$ and
a magnetic field with $\vaxd/\csd=0.29$.
Top: Run~B8, $\Ratio=0.099$; bottom: Run~B8h, $\Ratio=0.0098$.
}\label{ko_GSMa89}
\end{figure}

\subsection{$p\,$-modes}

The results of \cite{NT76} demonstrate that already for small values of the parameter
$\vA(0)/\cs$ there can be a noticeable influence on the eigenfrequencies
(see their Fig.~1).
However, it has to be considered that $\vA(z)/\cs$ grows to infinity due to the 
exponential decrease of density with height and the lack of an upper boundary
in their setup.
Hence, their findings cannot directly be transferred
to our model of finite thickness.   
Therefore we restrict ourselves to a comparison with
the non-magnetic case to infer the magnetically induced departures
of frequency, mode amplitude, mode mass, and line width.

\begin{figure*}
\begin{center}
\includegraphics[scale=1]{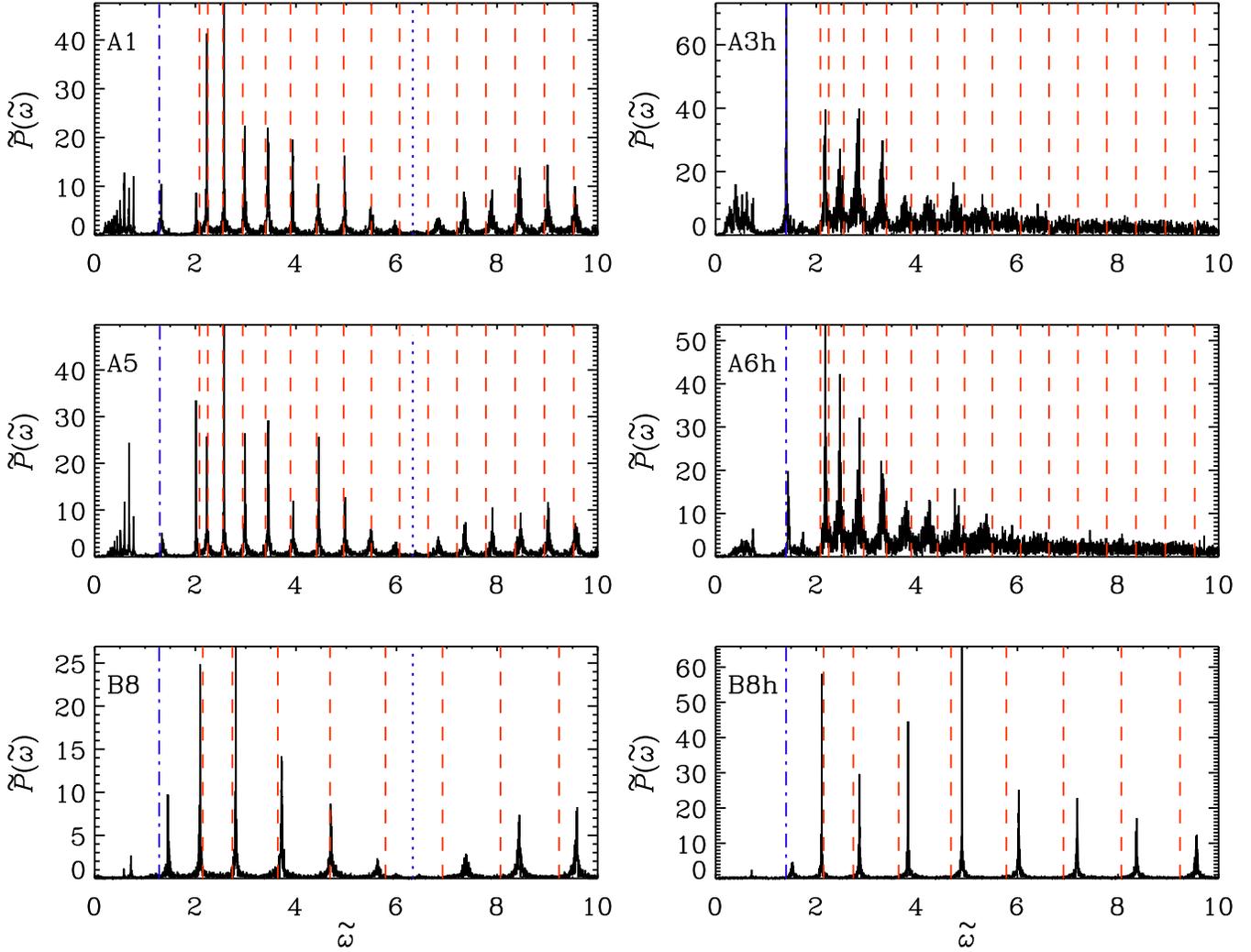}
\end{center}
\caption[]{$\widetilde{P}(\widetilde{\omega}; \widetilde{k}_x=2)$ 
as function of $\widetilde{\omega}$ with horizontal magnetic field for
Runs~A1 (a), A5 (c) and B8 (e), each with $\Ratio_0=0.1$,
whereas $\Ratio_0=0.01$ for Runs~A3h (b), A6h (d) and B8h (f);
see \Tab{table-SMH}.
Dash-dotted (blue) and dashed (red) lines:
theoretically expected locations of the $f$- and
$p\,$-modes, respectively.
Blue dotted lines: position of the $\omega=\csu k_x$ line,
shown dotted in the $k \omega$ diagrams (see e.g., \Fig{ko_TSMb2}).}
\label{fp-pow_kx2-HM}
\end{figure*}

In the $k\omega$ diagram, such as \Fig{ko_GSMa89}, we notice 
again the apparent gap in the $p\,$-mode spectrum coinciding with the 
separatrix $\omega=\csu k_x$ (dotted).
In \Fig{fp-pow_kx2-HM} we plot $\widetilde{P}$
as a function of $\widetilde{\omega}$ at $\widetilde{k}_x=2$ for six
cases with different field strengths, domain depths, and 
the two $\Ratio$ values, $0.1$ and $0.01$;
for details see \Tab{table-SMH}.
The dash-dotted blue lines show the theoretically expected locations of the 
non-magnetic $f$-mode; see \Eq{DR-NM}.
The group of peaks to the
left of the $f$-mode are the $g\,$-modes, whereas those 
to the right are $p\,$-modes.
The blue dotted lines mark the position of the separatrix
shown dotted in the $k\omega$ diagrams.
The red lines show the locations of theoretically expected
$p\,$-modes corresponding to the non-magnetic case;
see \Eqs{disp}{Quarter-Wave}.

Looking at different panels of \Fig{fp-pow_kx2-HM} and also
making some comparisons with the non-magnetic case in
\Fig{fp-pow_kx2-nM}, we notice that the frequencies of the individual
peaks of the $p\,$-modes are not much affected by the presence of
magnetic fields when $\Ratio_0=0.1$.
However, for the stronger density jump at the interface
($\Ratio_0=0.01$), slight frequency shifts of the $p\,$-modes may be
seen from the right panels of \Fig{fp-pow_kx2-HM}.
The mode amplitudes also seem to have increased compared to the corresponding
cases with $\Ratio_0=0.1$.

According to the work of \cite{HZ94},
albeit in the absence of a magnetic field,
the $p\,$-modes are expected to be strongly
affected by a hot outer corona.
Indeed, we see from our results that for the larger coronal temperature,
there is a noticeable effect on the $p\,$-modes,
which might not be connected with the magnetic field,
at least in the parameter regimes explored.   
Also, for higher coronal temperatures, the high frequency peaks
appear sharp in a shallower domain (Run~B8h), while
for the deeper domain (Runs~A3h and A6h) the data look more noisy.

\subsection{$f$-mode}
\label{fm_mode}

Given that the Alfv\'en speeds in the layers
above and below the interface are different due to 
the density jump at $z=0$, our setup 
mimics the `single magnetic interface' of \cite{R81}, \cite{MR89,MR92}, and
\cite{MAR92}.
It is capable of supporting a surface wave which, in the absence of gravity,
propagates with the phase speed $c_{\rm fm}$,
given by \citep{KS54,DL54,G67,MR89}
\EQ
c_{\rm fm}^2 = \frac{\ru \vaxu^2 + \rd \vaxd^2}{\ru+\rd} = 
\frac{2\ru \vaxu^2}{\rd+\ru}
=\frac{2 \rd \vaxd^2}{\rd+\ru};
\label{cm}
\EN
thus, $\vaxd \le c_{\rm fm} \le \vaxu$.
We note that this expression is valid for an incompressible fluid.
A more general result is given by \cite{R81}, but in our parameter
regime the difference is small.
The presence of gravity modifies the dispersion relation 
\citep{C61,R81,MR89,MR92,MAR92}:
\EQ
\omega_{\rm fm}^2 = c_{\rm fm}^2 k_x^2 +
g k_x \frac{1-\Ratio}{1+\Ratio},
\label{DR-SHM}
\EN
where $c_{\rm fm}^2 k_x^2$ always adds to the (squared) frequency
of the classical $f$-mode given by \Eq{DR-NM}. 

In \Figss{ko_GSMa1-2}{ko_GSMa89},
the dash-triple-dotted and dashed curves show the expected 
$f$-mode frequencies in the absence and presence of a horizontal
magnetic field, given by $\omega_{\rm f}$ and $\omega_{\rm fm}$,
respectively; see \Eqs{DR-NM}{DR-SHM}.
In \Fig{ko_GSMa1-2}, we show the $k\omega$ diagram for a case
with a weak horizontal magnetic field, $\vA/\cs=0.025$.
The frequencies of the $f$-mode are not yet noticeably affected.
However, when the field is increased by a factor of about five,
there is a clear frequency increase 
relative to $\omega_{\rm f}$ and we find
reasonably good agreement with $\omega_{\rm fm}$; see \Fig{ko_TSMb2}.

On the other hand, the amplitude of the $f$-mode diminishes
significantly for $\widetilde{k}_x\ga9$.
To investigate the reasons for the suppression of the $f$-mode 
at large horizontal wavenumbers, we performed more simulations
(not shown here) with much stronger hydrodynamic forcing.
We found that this did not significantly affect the
value of $k_x$ at which suppression apparently sets on,
suggesting that this cannot be a nonlinear effect.
From \Fig{EFs_p-f} we find that
the eigenfunctions of the $f$-modes, corresponding to non-magnetic
(open circles; Run~B) and horizontal magnetic field cases
(red filled circles; Run~B8), lie nearly on top of each other.
Thus arguments based solely on dissipative effects might
not be sufficient either to explain the mode suppression.
We therefore speculate that this effect could arise due to the inhibition of
vertical motions in the presence of a horizontal magnetic field
which is more pronounced for larger $k_x$.

In the case of a hotter corona ($\Ratio_0=0.01$), the
$f$-mode has larger amplitudes relative to the corresponding
case with the same magnetic field, but $\Ratio_0=0.1$,
and they extend up to $\widetilde{k}_x\approx 15$;
see \Fig{ko_GSMa4} for $\vA/\cs=0.042$ and compare with, for example,
\Fig{ko_GSMa1-2}, which is also for a weak (but different) field strength.
Again, however, significant frequency shifts can only be
seen for stronger magnetic fields; see \Fig{ko_TSMb3}
for $\vA/\cs=0.16$. Quantitative details and comparisons
are discussed later.

Following the procedure presented in \Sec{nM} to analyse the $f$-modes,
we determine the fit parameters 
at different values of $k_x$ for the cases in \Tab{table-SMH}.
Let us denote the numerical estimate of the line center
from the fit by $\ofmn$ and compute the relative frequency shifts as
\beq
\frac{\delta \omega_{\rm fm}^2}{\omega_{\rm f}^2} =
\frac{\oofmn-\omega_{\rm f}^2}{\omega_{\rm f}^2}.
\label{do2fm}
\eeq
In addition, we define the {\it theoretically expected} line shift due to the
magnetic field as
\beq
\left(\frac{\delta \omega_{\rm fm}^2}{\omega_{\rm f}^2}\!\right)_{\!\!\rm th} \equiv
\frac{\omega_{\rm fm}^2 - \omega_{\rm f}^2}{\omega_{\rm f}^2} =
\frac{2 \Ratio}{1-\Ratio}\, \frac{\vaxu^2 k_x}{g}.
\label{do2fm-th1}
\eeq
Note that $(\delta \omega_{\rm fm}^2/\omega_{\rm f}^2)_{\rm th}$
is, for $\Ratio \ll 1$, proportional to $\Ratio$
so it increases with decreasing density contrast.
Furthermore, $(\delta \omega_{\rm fm}^2/\omega_{\rm f}^2)_{\rm th}$
increases also with $k_x$, so it should become
more noticeable at small length scales.
Expressing it in terms of the Alfv\'en speed in the bulk, we have    
\beq
\left(\frac{\delta \omega_{\rm fm}^2}{\omega_{\rm f}^2}\!\right)_{\!\!\rm th} =
\frac{2}{1-\Ratio}\; \frac{\vaxd^2 k_x}{g} =
\frac{2}{1-\Ratio}\; \frac{\vaxd^2}{\csd^2}\;
\widetilde{k}_x.
\label{do2fm-th2}
\eeq
For the definition of mode mass $\mu_{\rm f}$ and
linewidth $\Gamma_{\rm f}$, see \Eqs{muf}{lwf}. 

\begin{figure}
\begin{center}
\includegraphics[width=\columnwidth]{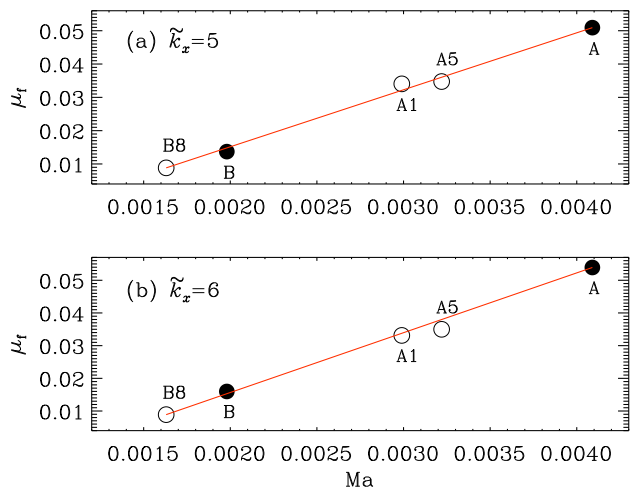}
\end{center}
\caption[]{Mode mass of the $f$-mode as a function of $\md$ for
$\Ratio_0=0.1$ and two values of $\widetilde{k}_x$
(see \Tabs{table-SnM}{table-SMH}).
Symbols: values from DNS.
Red lines: linear fits with slopes $\sim 16.9$ (a)
and $\sim 18.3$ (b).
}
\label{fmm-Md}
\end{figure}

We have already noted that an increase of the magnetic field
diminishes the $f$-mode mass already for smaller values of $k_x$.
To understand this quantitatively, we compare with the
non-magnetic case, where we have seen a decrease of the
mode mass with decreasing Mach number.
It turns out that the magnetic effect on the mode mass
can be understood solely as a consequence of the reduction of $\Ma$,
for a given $\Ratio_0$.
This is shown in \Fig{fmm-Md}, where we plot $\mu_{\rm f}$
versus $\md$ for magnetic and non-magnetic cases:
both sets 
exhibit the same $\md$ dependence.

\begin{figure*}
\begin{center}
\includegraphics[scale=1]{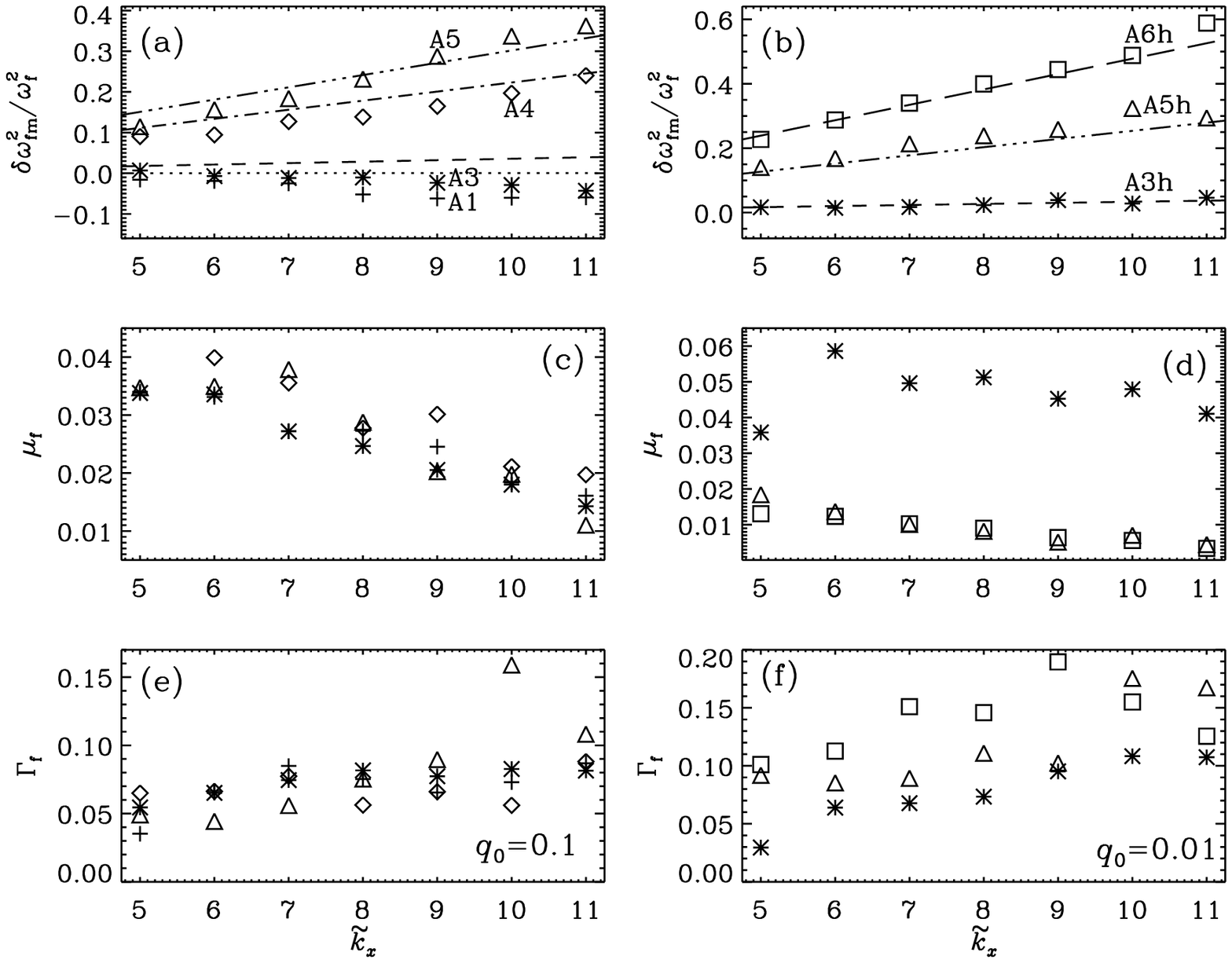}
\end{center}
\caption[]{Properties of the $f$-mode as functions of $\widetilde{k}_x$,
for $\Ratio_0=0.1$ (left) and
$\Ratio_0=0.01$ (right). 
Lines in panels (a) and (b):
theoretical estimates from \eqref{do2fm-th1}.
The `$+$' (dotted), `$\ast$' (dashed),
`$\diamond$' (dash-dotted) and `$\triangle$' (dash-triple-dotted) in
panels (a), (c) and (e) correspond to
Runs~A1, A3, A4 and A5, respectively.
The `$\ast$' (dashed), `$\triangle$' (dash-triple-dotted) and
`$\square$' (long-dashed) in panels (b),
(d) and (f), respectively, correspond to Runs~A3h, A5h and A6h, respectively.}
\label{fmode-kx_HM}
\end{figure*}

\begin{figure}
\begin{center}
\includegraphics[width=\columnwidth]{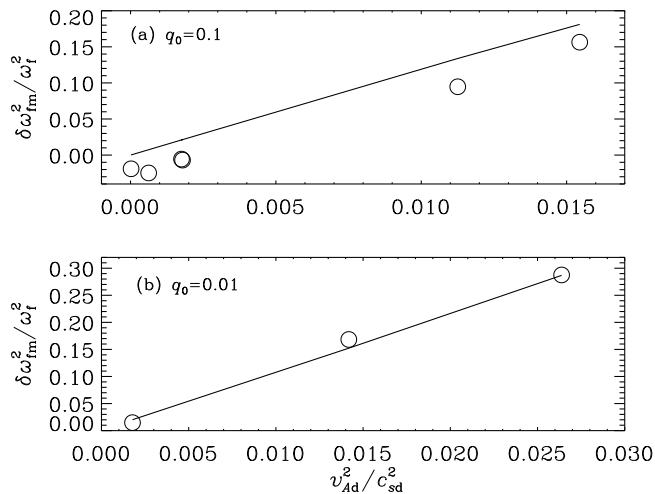}
\end{center}
\caption[]{
Relative frequency shift $\delta \omega_{\rm fm}^2/\omega_{\rm f}^2$
as a function of 
$\vad^2/\csd^2$
at $\widetilde{k}_x=6$.
Lines: estimates from \eqref{do2fm-th2}, symbols: values from DNS.
}\label{fshift-vA2}
\end{figure}

The dependence of $\delta\omega_{\rm fm}^2/\omega_{\rm f}^2$,
$\mu_{\rm f}$ and $\Gamma_{\rm f}$
on the horizontal wavenumber $k_x$ 
in the presence of a magnetic field
is shown in \Fig{fmode-kx_HM} for some of the cases of \Tab{table-SMH},
covering both $\Ratio_0=0.1$ and $\Ratio_0=0.01$. 
We recall that, while magnetic fields tend to increase the $f$-mode
frequencies \citep{C61,R81,MR92,MAR92},
turbulent motions lead to a decrease
\cite{M00a,M00b}; see \Sec{fmode_nonmag}.
For weak magnetic fields (Runs~A1 and A3 with $\vaxd/\csd=0.004$
and $0.042$, respectively)
we find at large values of $\widetilde{k}_x$ considerable
frequency decrements compared to the theoretical
estimates when $\Ratio_0=0.1$; see  panel (a) of \Fig{fmode-kx_HM}.
This is because the magnetic field might here not be sufficient
to compensate for the decrease caused by the turbulence.
As the strength of the imposed field is increased,
$\omega_{{\rm fm}\#}$ also increases and shows reasonably
good agreement with the theoretically expected values
\eqref{DR-SHM} for Runs~A4 and A5.
For the hotter corona with $\Ratio_0=0.01$, numerical findings and 
theoretical estimates lie even nearly on top of each other,
see panel (b).
Depending on the strength of the field,
the frequencies can be much higher than those of the
classical $f$-mode given by \Eq{DR-NM}.
It is interesting to note that,
although $\vA/\cs=0.042$ for both A3 and A3h,
the relative frequency shifts are different and smaller
for the stronger density jump; compare
panels (a) and (b) of \Fig{fmode-kx_HM}.
From panels (c)--(f) we note:
\begin{enumerate}
\item [\phantom{ii}(i)] \parbox[t]{.877\linewidth}{
The mode mass $\mu_{\rm f}$ decreases with $k_x$
for all runs. It decreases more rapidly when 
the density jump is smaller.}\\[0mm]
\item [\phantom{i}(ii)] \parbox[t]{.877\linewidth}{
The distinguishing effect of the imposed field
is that the $f$-mode is suppressed
beyond some $\kfm$, which decreases as we increase $B_{x0}$; 
compare, e.g., the $k\omega$ diagrams
\Fig{ko_GSnMa1} with \Fig{ko_TSMb2} or
\Fig{ko_GSnMa1_8pixpi} with \Fig{ko_GSMa89}a, 
for $\Ratio_0=0.1$ and \Fig{ko_GSMa4} with 
\Fig{ko_TSMb3} for $\Ratio_0=0.01$, respectively.}\\[0mm]
\item [(iii)] \parbox[t]{.877\linewidth}{
For $\Ratio_0=0.1$, $\mu_{\rm f}$ does not show any
systematic variation with $B_{x0}$ at a fixed $k_x$;
see panel (c).
But for $\Ratio_0=0.01$, $\mu_{\rm f}$ decreases drastically
with increasing $B_{x0}$ for all $k_x$;
see panel (d).}\\[0mm]
\item [(iv)] \parbox[t]{.877\linewidth}{
For  small $B_{x0}$, $\mu_{\rm f}$ is larger 
for $\Ratio_0=0.01$ compared to $\Ratio_0=0.1$;
compare runs~A3 and A3h, for both
of which $\vA/\cs=0.042$.}\\[0mm]
\item [\phantom{i}(v)] \parbox[t]{.877\linewidth}{
For large $B_{x0}$, $\kfm$ is smaller for $\Ratio_0=0.01$ 
than for $\Ratio_0=0.1$; compare both panels of \Fig{ko_GSMa89}.
This is the reason why 
$\mu_{\rm f}$ of run A5 is, for small $\widetilde{k}_x$,
larger compared to run A5h (both with $\vA/\cs\approx 0.12$),
despite the latter having stronger density contrast.}\\[0mm]
\item [(vi)] \parbox[t]{.877\linewidth}{
For most runs, the line width $\Gamma_{\rm f}$ increases with
$k_x$; see panels (e) and (f).
Although it does not show any systematic variation with $B_{x0}$ for
fixed $k_x$ when $\Ratio_0=0.1$, it increases with $B_{x0}$
at all $k_x$ when $\Ratio_0=0.01$.}
\end{enumerate}
To quantify the magnetically produced line shifts, we now consider Runs~A1--A6h.
It turns out that $\delta \omega_{\rm fm}^2/\omega_{\rm f}^2$
is approximately proportional to $\vaxd^2/\csd^2$,
as expected from theory; see \Fig{fshift-vA2} and
\Eq{do2fm-th2}. We find somewhat larger frequency shifts
compared to theory when $\Ratio_0=0.1$, but better agreement
when $\Ratio_0=0.01$.

\begin{figure}
\begin{center}
\includegraphics[width=\columnwidth]{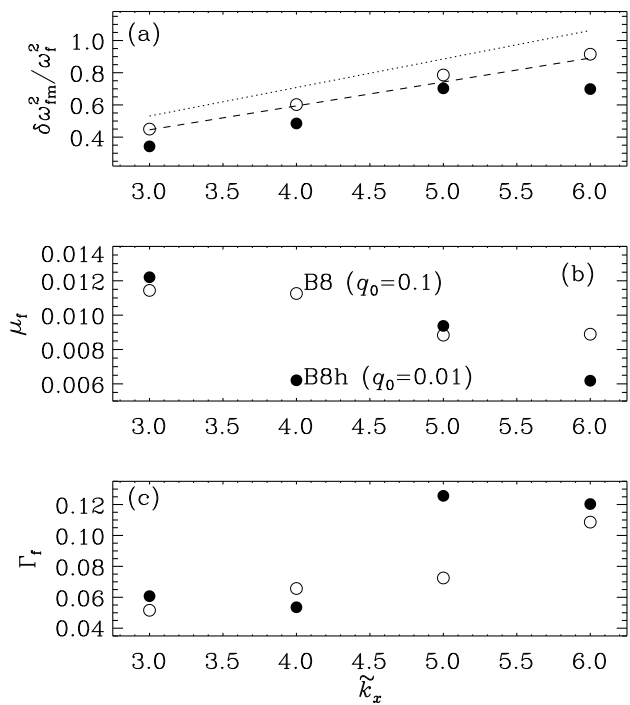}
\end{center}
\caption[]{
Effect of hot corona on mode parameters for horizontal magnetic field. 
Open/filled circles: Run~B8/B8h, both with $\vA/\cs \approx 0.3$.
Dotted/dashed lines in upper panel:  theoretical estimates for B8/B8h.}
\label{fmode-CorEff}
\end{figure}

To assess the effect of a hotter corona, we plot in
\Fig{fmode-CorEff} the $k_x$ dependencies of 
$\delta\omega_{\rm fm}^2/\omega_{\rm f}^2$,
$\mu_{\rm f}$ and $\Gamma_{\rm f}$
for models B8 ($\Ratio_0=0.1$) and B8h ($\Ratio_0=0.01$).
Both runs are for a shallower domain with $L_z=\pi$,
and have $\vaxd/\csd \approx 0.3$; see \Tab{table-SMH}.
Interestingly, the line shifts are slightly reduced
for a hotter corona.
This behaviour is in agreement with the $\Ratio$
dependence of the shifts expected from the theoretical 
result \eqref{do2fm-th1}.
However,
the numerically obtained shifts lie below the theoretical estimates for both runs.
As the imposed field $B_{x0}$ is large, the $f$-mode 
is truncated beyond relatively small values of $\widetilde{k}_x$, 
as may be inferred from \Fig{ko_GSMa89}. 
For strong magnetic fields, the Mach numbers are small despite 
strong forcing; see \Tab{table-SMH}.
Consequently, the mode mass $\mu_{\rm f}$ is small in both runs, but it
drops more rapidly with $k_x$ for $\Ratio_0=0.01$ than for
$\Ratio_0=0.1$; compare panels
(c) and (d) of \Fig{fmode-kx_HM} with panel (b) of \Fig{fmode-CorEff}
(note the different ranges of the $\widetilde{k}_x$ axes).
From panel (c) of \Fig{fmode-CorEff} we note that
the line width $\Gamma_{\rm f}$ increases with $k_x$ and
is larger for smaller $\Ratio_0$.
All these observations are in qualitative agreement with what we noted before
in case of deeper domains; see \Fig{fmode-kx_HM}.

\subsection{$g\,$-modes}

Remarkably, the $g\,$-modes are strongly suppressed beyond
some value $k_x^{g\rm max}$, which is found to decrease 
with increasing $B_{x0}$;
compare, e.g., \Fig{ko_GSnMa1} with \Figs{ko_GSMa1-2}{ko_TSMb2},
or \Fig{ko_GSnMa1_8pixpi} with \Fig{ko_GSMa89}(a).
Varying $\Ratio_0$ for fixed $B_{x0}$ does not seem
to have much effect on the $g\,$-modes;
cf.\ Figs.~\ref{ko_GSMa89}(a) and (b).

\begin{table*}\caption{
Summary of simulations with a vertical and oblique
magnetic fields in domain $8\pi \times 2\pi$. $\theta$ -- inclination of $\BB_0$ to $z$ axis.
}\centering
\label{table-SMV}
\begin{tabular}{l c c c c c c c c c c c c c}
\hline\hline
Run & $\theta$ & Grid & $\Ratio_0$ & $\Ratio$ & $\vad/\csd$ & $\vau/\csu$ & $\md$ & $\re$ & ${\cal F}$\\ [-0.2ex]
\hline
A2V & $0^\circ$ & $1024\times 600$ & 0.1 & 0.087 & 0.020 & 0.020 & 0.0030 & 0.536 & 0.02\\
\hline
A6Vh &$0^\circ$ &  $1024\times 512$ & 0.01 & 0.0089 & 0.117 & 0.118 & 0.0011 & 0.256 & 0.02 & (\Fig{ko_TSMb1-vert})\\
A7Vh & $0^\circ$ &$1024\times 512$ & 0.01 & 0.0096 & 0.185 & 0.176 & 0.0016 & 0.730 & 0.04\\
Obl  &  $45^\circ$ &$1024\times 512$  & 0.01  & 0.009 & 0.22 & 0.22 & 0.0015 & 0.35 & 0.02 &  \\
\hline
\end{tabular}
\end{table*}

\section{Vertical and oblique fields}
\label{GM}

We now turn to cases where the imposed magnetic
field is either vertical or points in 
an oblique direction in the $xz$ plane.
There have been earlier attempts to study the interaction of the
$f$- and $p\,$-modes with a vertical magnetic field to provide
some explanation for the observed absorption of these modes
\citep{CB93,CBZ94,CB97}.
Although the physics of
mode absorption is yet to be understood, some explanations
in terms of {\em slow mode leakage} due to vertical stratification
were discussed in these works:
it is thought that the
partial conversion of the $f$- and $p\,$-modes
to slow magnetoacoustic modes, or $s$-modes, takes place whenever
they encounter
the region of a vertical magnetic field, such as a sunspot.
\cite{PK09} numerically investigated the effects of an inclined
magnetic field on the excitation and propagation of helioseismically relevant
magnetohydrodynamic waves.
They found that the
$f$-modes are affected by the background magnetic field more than the
$p\,$-modes.
Such results emphasize the diagnostic role of $f$-modes to reveal the
subsurface structure of the magnetic field.

\begin{figure}
\begin{center}
\includegraphics[width=\columnwidth]{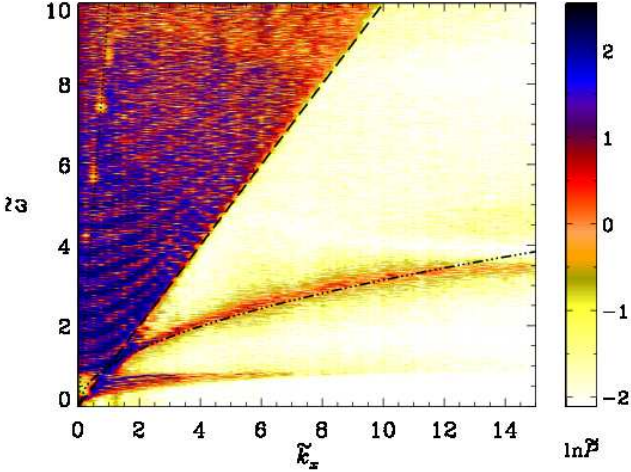}
\end{center}
\caption[]{$k\omega$ diagram for Run~A6Vh with vertical magnetic field;
see \Tab{table-SMV}.
Dash-triple-dotted: $\omega_{\rm f}$ as defined in \Eq{DR-NM}.
Long dashed: $\omega=\csd k_x$.
}\label{ko_TSMb1-vert}
\end{figure}

In \Tab{table-SMV} we summarize the simulations with vertical
and oblique magnetic fields. 
With a vertical magnetic field $(0,0,B_{z0})$, the Alfv\'en velocities
$v_{{\rm A}z\,{\rm d,u}}$ in bulk and corona are defined 
analogously to \Eq{vAx-ud}.

In \Fig{ko_TSMb1-vert} we present the $k\omega$ diagram for Run~A6Vh.
Remarkably, the frequency shift of the $f$-mode shows a non-monotonic
behaviour, unlike those in other $k\omega$ diagrams.
Here the frequencies lie above those
of the classical $f$-mode given by \Eq{DR-NM}
(dash-triple-dotted curves in $k\omega$ diagrams)
for intermediate wavenumbers, while 
falling below it
for larger wavenumbers.
Interestingly, both tendencies have been reported in observational
results of \cite{FSTT92}.

We report on one simulation where the magnetic 
field points in a direction of
$45^{\circ}$ to the $z$-axis.
It was performed with $\Ratio_0=0.01$, 
and $\vaxd/\csd=\vazd/\csd=0.157$,
or $\vA/\cs=0.22$ for the total field
as $\vA^2=v_{{\rm A}x}^2+v_{{\rm A}z}^2$.

\subsection{$p\,$-modes}

\begin{figure*}
\begin{center}
\includegraphics[scale=1]{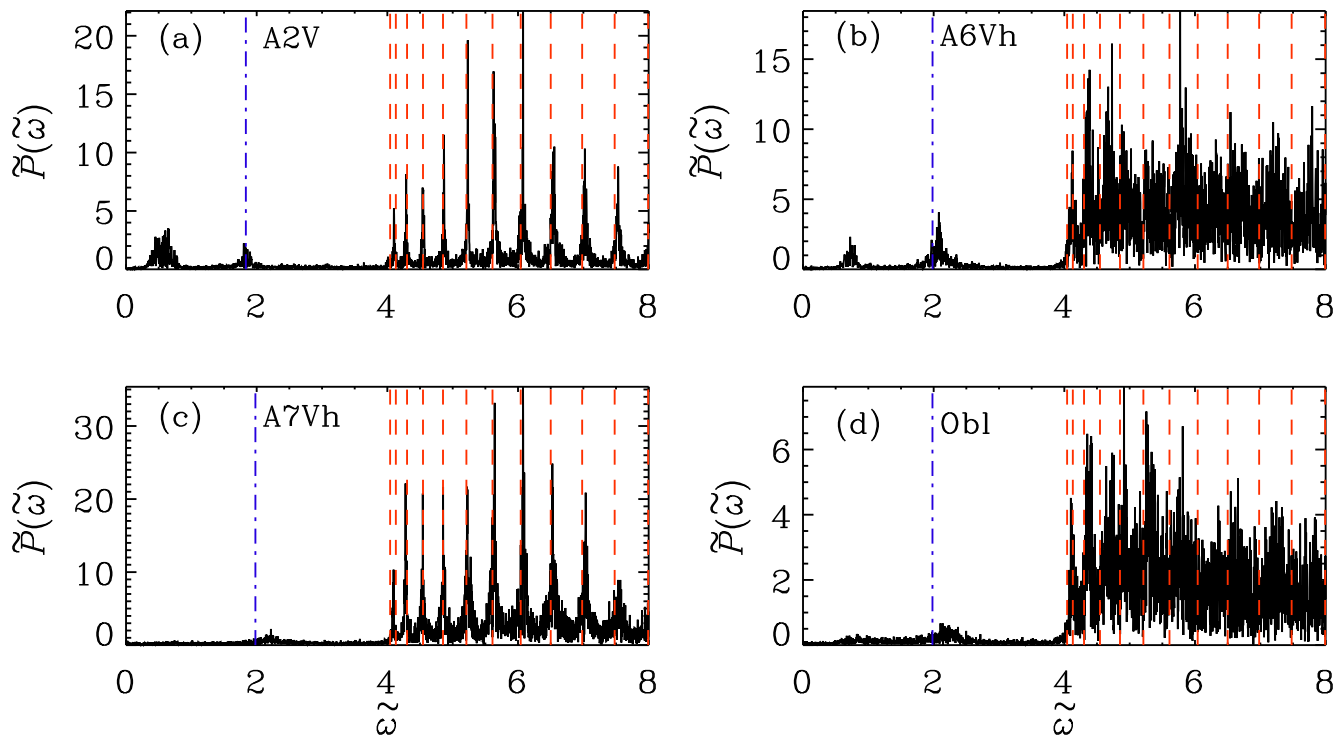}
\end{center}
\caption[]{$\widetilde{P}(\widetilde{\omega}; \widetilde{k}_x=4)$ 
for a vertical magnetic field for (a) A2V, (b) A6Vh, (c) A7Vh, and
(d) for the $45^\circ$ inclined magnetic field;
see \Tab{table-SMV}.
Dash-dotted (blue) and dashed (red):
theoretically expected $f$- and
$p\,$-modes, respectively, all for the non-magnetic case. 
}\label{fp-pow-Bgen-kx4}
\end{figure*}

In \Fig{fp-pow-Bgen-kx4} we plot $\widetilde{P}(\widetilde{\omega})$
at $\widetilde{k}_x=4$ for three runs with
vertical field (A2V, A6Vh, A7Vh), 
and for the one with oblique field (Obl);
see also \Tab{table-SMV}.
As before, the dash-dotted (blue) and dashed (red) lines in all panels
mark the theoretically expected locations of the $f$- and
$p\,$-modes, respectively, all for the non-magnetic case. 
The group of peaks to the left of the $f$-mode are the $g\,$-modes,
whereas those to the right are $p\,$-modes.
Note that for higher coronal temperatures, the modes are much more
noisy and there appears to be a larger continuum; 
cf. panels (a) and (d) of \Fig{fp-pow-Bgen-kx4}.
For weaker jumps in the thermodynamic quantities at the
interface, we find that $p\,$-mode amplitudes
are not much affected by the presence of a weak vertical field;
compare e.g. Runs~A (non-magnetic) and A2V shown in
\Figs{fp-pow_kx4-nM}{fp-pow-Bgen-kx4}, respectively,
both having $\Ratio_0=0.1$.
Compared to other cases, we notice a significant
reduction in the $p\,$-mode amplitudes
in case of the inclined magnetic field, which has $\Ratio_0=0.01$;
see panel (d).
The $g\,$- and $f$-modes are also strongly suppressed,
which might be caused by the large
strength of the magnetic field, leading to the truncation of
these modes beyond some wavenumber, $k_x\approx 4$.

\subsection{$f$-mode}

\begin{figure}
\begin{center}
\includegraphics[width=\columnwidth]{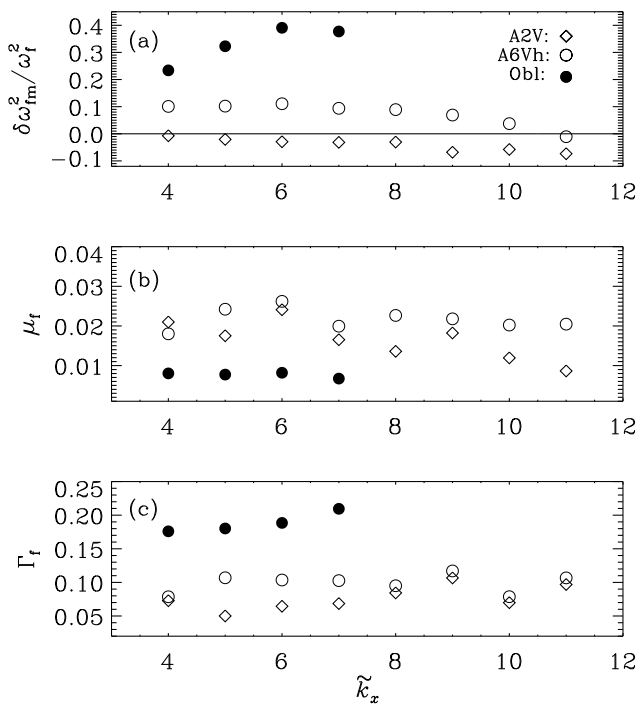}
\end{center}
\caption[]{(a) Relative frequency shift, (b) mode mass and (c) linewidth of
the $f$-mode as  functions of $\widetilde{k}_x$ in the presence of a
vertical/oblique magnetic field.
Diamonds and open circles:
runs A2V and A6Vh, respectively, with vertical field;
filled circles: with $45^\circ$ inclined 
magnetic field; all with comparable Mach numbers.
}\label{fmode-kx_GM}
\end{figure}

An analytical dispersion relation of the $f$-mode 
for vertical magnetic field suitable to our setup
is yet to be derived.
For quantitative analysis, we compute 
relative frequency shift
$\delta \omega_{\rm fm}^2/\omega_{\rm f}^2$, mode mass $\mu_{\rm f}$,
and dimensionless line width $\Gamma_{\rm f}$
of the $f$-mode for different $k_x$,
following the procedures described in \Sec{nM}.
In \Fig{fmode-kx_GM} we show the dependence of the line parameters
on $k_x$ for Runs~A2V (with $\Ratio_0=0.1$),
A6Vh and Run~Obl (with $\Ratio_0=0.01$ for the
latter two runs).
Some noteworthy observations are:
\begin{enumerate}
\item [\phantom{ii}(i)] \parbox[t]{.877\linewidth}{The relative line shift
$\delta \omega_{\rm fm}^2/\omega_{\rm f}^2$
shows a non-monotonic behaviour as
a function of $k_x$ for sufficiently strong 
$B_{z0}$, unlike our findings of
\Secs{nM}{HM} without field or with horizontal field.
This holds also for Run~A6Vh where
we find that $\delta \omega_{\rm fm}^2/\omega_{\rm f}^2$
reaches a maximum at $\widetilde{k}_x=6$,
and becomes negative 
at $\widetilde{k}_x=11$. }\\[0mm]
\item [\phantom{i}(ii)] \parbox[t]{.877\linewidth}{For weak field, 
$\delta \omega_{\rm fm}^2/\omega_{\rm f}^2$
is negative and decreases with increasing $k_x$ as 
in non-magnetic cases.}\\[0mm]
\item [(iii)] \parbox[t]{.877\linewidth}{For large $B_{z0}$,
with or without $B_{x0}$, we find that
$\delta \omega_{\rm fm}^2/\omega_{\rm f}^2$
attains positive values for small $\widetilde{k}_x$.
}\\[0mm]
\item [(iv)] \parbox[t]{.877\linewidth}{
For Run~Obl, we notice larger positive 
frequency shifts compared to Run~A6Vh.
It increases up to about
$\widetilde{k}_x\approx 6$, beyond which it tends to decrease.
Fewer points (filled circles) are shown as the $f$-mode is
truncated for larger $\widetilde{k}_x$ due to 
the stronger magnetic field
compared to the other runs shown. This truncation effect is
also discussed earlier.}\\[0mm]
\item [\phantom{i}(v)] \parbox[t]{.877\linewidth}{The mode masses 
from Runs~A2V and A6Vh are comparable, despite the latter
having smaller Mach number;
see \Tab{table-SMV}. This is due to the stronger density
jump at the interface in Run~A6Vh,
and thus consistent with our earlier
findings.}\\[0mm]
\item [(vi)] \parbox[t]{.877\linewidth}{As the 
field strength is increased in Run~Obl compared with Run~A6Vh, 
$\mu_{\rm f}$ decreases at all $k_x$
although both runs have similar Mach numbers.}\\[0mm]
\item [(vii)] \parbox[t]{.9\linewidth}{The line width $\Gamma_{\rm f}$ increases with
$k_x$ and is larger for stronger fields.}
\item [(viii)] \parbox[t]{.9\linewidth}{Compared to the horizontal magnetic
field cases, the $f$-mode suppression is not seen for large values of $k_x$;
cf.\ \Figs{ko_TSMb3}{ko_TSMb1-vert}.
For vertical magnetic fields, the energy could in principle leave the
interface, leading to a reduction of $f$-mode power.
This however does not apply in our case owing to the perfect
conductor boundary condition.}
\end{enumerate}

\subsection{$g\,$-modes}

The $g\,$-modes are found to behave similarly as
with horizontal magnetic field and are suppressed beyond some
$k_x^{g\rm max}$, which 
decreases
with increasing field.
This may be inferred from \Fig{fp-pow-Bgen-kx4} where the
$g\,$-modes are seen to be suppressed at $\widetilde{k}_x=4$
with increasing field for both the vertical and the oblique field;
compare also with the non-magnetic case in \Fig{fp-pow_kx4-nM}.

\section{Conclusions}

The prime objective of the present work was to assess the effects
of an imposed magnetic field on the $f$-mode, which is known to be
particularly sensitive to magnetic fields.
One of our motivations is the ultimate application to cases where
magnetic flux concentrations are being produced self-consistently
through turbulence effects \citep[see, e.g.,][]{BKR13,BGJKR14}.
Those investigations have so far mostly been carried out in isothermal
domains, which was also the reason for us to choose a piecewise
isothermal model, where the jump in temperature and density is
needed to allow the $f$-mode to occur.

The resulting setup is in some respects different from that of the Sun
and other stars, so one should not be surprised to see features
that are not commonly found in the context of helioseismology.
One of them is a separatrix within the $p\,$-modes
as a result of the hot corona, which we associate with the
$a$-mode of \cite{HZ94}.

Regarding the $f$-mode, there are various aspects that can be
studied even in the absence of a magnetic field.
Particularly important is the linearly $k_x$-dependent decrease
of $\omega_{\rm f}$.
The $f$-mode mass increases with the intensity of the forcing and
hence with the Mach number.
Interestingly, this is also a feature that carries over to the
magnetic case where an increase in the magnetic field leads to
a decrease in the resulting Mach number and thereby to a decrease
in the mode mass in much the same way as in the non-magnetic case.
Magnetic fields also lead to a truncated $f$-mode branch above
a certain value of $k_x$.

One of the most important findings is the systematic increase of
the $f$-mode frequencies $\ofmn$ observed in DNS with the horizontal
magnetic field.
It follows essentially the theoretical prediction
and, contrary to the non-magnetic cases, shows an increase with
$k_x$ such that the relative frequency shift is approximately
proportional to $\vA^2/\cs^2$.
This is best measured when $k_x L_0=5$--$7$,
i.e., $k_x\Hp=3$--$4$, and
with a solar radius of $700\Mm$, being 2000 times larger than
$\Hp\approx300\kmeter$, the corresponding spherical harmonic degree
would be 6000--8000.
In this range, the relative frequency shift is
$\delta\omega_{\rm fm}^2/\omega_{\rm f}^2\approx0.1$
when $\vA/\cs\approx0.1$.
Since $\delta\omega_{\rm fm}^2/\omega_{\rm f}^2 \approx 2\delta\omega_{\rm fm}/\omega_{\rm f}$,
the increase of the $f$-mode 
frequencies
is about 5\%.
The observed $f$-mode frequency increase during solar maximum
is about $1\uHz$ at a moderate spherical degree of 200 \citep{DG05}.
This corresponds to a relative shift of about 0.06\%,
but this value should of course increase with the spherical degree.

We note that in our case, the magnetic field is the same above
and below the interface.
Furthermore, $\rho\cs^2$ is also the same above and below the interface,
and so is therefore also $\vA^2/\cs^2$.
One of the goals of future studies will be to determine how our results
would change if the magnetic field existed only below the interface.

Interestingly, for vertical and oblique magnetic fields, the
$k_x$-dependence of $\ofmn$ becomes non-monotonic in such a way that
for small values of $k_x$, $\ofmn$ first increases with $k_x$ and then
decreases and becomes less than the theoretical value
in the absence of a magnetic field, although it stays above the 
value obtained without magnetic field, whose reduction
is believed to be due to turbulence effects \citep{MMR99,MKE08}.

We confirm the numerical results of \cite{PK09} that $p\,$-modes are
less affected by a background magnetic field than the $f$-mode.
Relative to the non-magnetic case, no significant frequency shifts
of $p\,$-modes are seen in a weakly magnetized environment.
For a larger density contrast at the interface,
with the rest of the parameters being same,
the mode amplitudes and line widths increase, but the data look
more noisy and the frequency shifts, which can be of either sign
\citep{HZ94}, may not be primarily due to the magnetic field.

The present investigations allow us now to proceed to more complicated
systems where the magnetic field shows local flux concentrations 
which might ultimately resemble active regions and sunspots.
As an intermediate step, one could also impose a non-uniform magnetic
field with a sinusoidal variation in the horizontal direction.
This will be the focus of a future investigation.

\section*{Acknowledgements}

We thank Aaron Birch for useful comments on an earlier version
of the manuscript and pointing out problems with insufficient cadence.
S.M.C. thanks Nordita for hospitality during an extended visit
in 2013 when this work began.
Financial support from the Swedish Research Council under the grants
621-2011-5076 and 2012-5797, the European Research Council under the
AstroDyn Research Project 227952 as well as the Research Council of
Norway under the FRINATEK grant 231444 are gratefully acknowledged.
The computations have been carried out at the National Supercomputer
Centres in Link\"oping and Ume{\aa} as well as the Center for Parallel
Computers at the Royal Institute of Technology in Sweden and the Nordic
High Performance Computing Center in Iceland.

%r e f

\end{document}